\documentclass[%
 aip,
 amsmath,amssymb,
preprint,%
]{revtex4-1}

\DeclareSymbolFont{newfont}{OML}{cmm}{m}{it}
\DeclareMathSymbol{\epsilon}{3}{newfont}{15}

\usepackage{graphicx}
\usepackage{dcolumn}
\usepackage{bm}
\usepackage{float}
\usepackage[utf8]{inputenc}
\usepackage[T1]{fontenc}
\usepackage{mathptmx}
\usepackage{multirow}
\usepackage{subcaption}
\usepackage{xcolor}

\begin{document}


\title{Thermal Transport Dynamics in Active Heat Transfer Fluids (AHTF)}

\author{Wei Peng}
\affiliation{Department of Materials Science and Engineering, Rensselaer Polytechnic Institute, Troy, New York 12180, USA}%

\author{Anirban Chandra}
\affiliation{Department of Mechanical, Aerospace, and Nuclear Engineering, Rensselaer Polytechnic Institute, Troy, New York 12180, USA}%

\author{Pawel Keblinski}
\affiliation{Department of Materials Science and Engineering, Rensselaer Polytechnic Institute, Troy, New York 12180, USA}%

\author{Jeffrey L. Moran}
\email{jmoran23@gmu.edu}
\affiliation{Department of Mechanical Engineering, George Mason University, Fairfax, VA 22030, USA}

\date{\today}

\begin{abstract}
We present results of molecular dynamics (MD) calculations of the effective thermal conductivity of nanofluids containing self-propelled nanoparticles. The translational and rotational dynamics observed in the simulations follow the behavior expected from the standard theoretical analysis of Brownian and self-propelled nanoparticles. The superposition of self-propulsion and rotational Brownian motion causes the behavior of the self-propelled nanoparticles to resemble Brownian diffusion with an effective coefficient that is larger than the standard Brownian value by a factor of several thousand. As a result of the enhanced diffusion (and the convective mixing resulting from the motion), we observe a discriminable increase of the effective thermal conductivity of the solution. While the increases we observe are in the range of several percent, they are significant considering that, without propulsion, the nanofluid thermal conductivity is essentially not affected by the Brownian motion and can be understood within the effective medium theory of thermal conduction. Our results constitute a proof of concept that self-propelled particles have the potential to enhance thermal conductivity of the liquid in which they are immersed, an idea that could ultimately be implemented in a broad variety of cooling applications.
\end{abstract}

\maketitle
\newcommand{\etal}{\textit{et al. }}

\section{Introduction}
Heat transfer fluids (HTF) are critical components of many engineering systems in which materials or devices need to be cooled, heated, or kept within a certain temperature range. HTF absorb, transport, and expel heat in applications such as solid-state lighting, automobile thermal management (e.g. brake fluid), high-power radio-frequency devices, and microelectronics cooling. Many of these technologies face fundamental performance limits that are determined in part by waste heat removal. The efficacy of HTF, and thus the performance of these devices, depend primarily on flow conditions, flow geometry, and fluid properties such as thermal conductivity, viscosity, and specific heat. Recent efforts have been especially devoted to increasing the thermal conductivity of liquids.

In the 1990s, Choi and Eastman introduced the first nanofluids, which are colloidal suspensions of nanoparticles in liquids \cite{choi1995enhancing}. Since solids generally conduct heat more efficiently than liquids, the nanoparticle suspension typically exhibits a higher thermal conductivity than the liquid alone. Soon after nanofluids were introduced, it was reported that nanofluids exhibited unusual thermal properties. These included thermal conductivity enhancements exceeding those predicted by classical theories, such as the effective-medium theory formulated by Maxwell in the 1880s \cite{maxwell1873treatise}; (2) a nonlinear dependence of thermal conductivity enhancement on nanoparticle volume fraction; (3) a dependence of thermal conductivity enhancement on nanoparticle size and shape; (4) a dependence of thermal conductivity enhancement on liquid temperature. 

The first observation—anomalous thermal conductivity enhancement—received special attention, and many theories were initially proposed to explain it. For example, it was hypothesized that `micro-convection' generated by Brownian motion of the particles contributed an additional mode of heat transfer, which might potentially account for the anomalous measurements \cite{prasher2005thermal}. However, this hypothesis is implausible considering that the nanoparticle diffusivity is generally orders of magnitude smaller than the thermal diffusivity of the base liquid. The latter observation is consistent with molecular dynamics (MD) simulations of thermal transport in nanofluids \cite{evans2006role, babaei2013proof}, as well as carefully-executed experiments on nanofluids with well-dispersed nanoparticles \cite{eapen2007mean, putnam2006thermal}. In 2009, researchers from 34 organizations worldwide conducted careful measurements of the thermal conductivity of various nanofluids and found that the effective-medium theory formulated by Maxwell \cite{maxwell1873treatise} (generalized by Nan et al. \cite{nan1997effective}), which ignores micro-convection effects, successfully predicted the thermal conductivity of the nanofluids tested to within experimental uncertainty, suggesting that no anomalous enhancement of thermal conductivity was evident in these nanofluids \cite{buongiorno2009benchmark}. Today, Brownian-motion-induced micro-convection is understood to play a negligible role in the transport of thermal energy through nanofluids \cite{babaei2013proof}.

Most anomalous measurements of nanofluid thermal conductivity are understood to originate from inconsistent measurement techniques and conditions and/or from nanoparticle agglomeration. The extent of agglomeration in nanofluids can depend on synthesis and dispersion conditions, and it can also change with time. In this context, the anomalous thermal conductivity enhancements can be attributed to the formation of sparse and fractal nanoparticle agglomerates that provide an interconnected conductive path for heat transport. However, such agglomerates tend to dramatically increase the suspension’s viscosity. Significant viscosity increases harm HTF performance by increasing the required pumping power and suppressing natural convection. The heat transfer enhancement potential of traditional nanofluids is thus limited. Nevertheless, as devices are further miniaturized, and power dissipation requirements simultaneously rise, the need for high-performance heat transfer fluids continues to increase rapidly. 

In the past 15 years there has been significant interest in colloidal particles that propel themselves in liquids. These ``active colloids'' harvest energy from their surroundings and convert it to their own kinetic energy through a variety of physico-chemical mechanisms. Synthetic active colloids range in size from $\sim 30 nm$ \cite{lee2014self} to $\sim 20\mu m$ \cite{gao2012water} and obtain energy for their propulsion from a variety of sources, such as chemical fuels, ultraviolet light, electric or magnetic fields, or ultrasound. 

As they move, both natural and synthetic microswimmers agitate the fluid around them, causing disturbance flows in their vicinity. Microswimmers are often classified as `pushers' or `pullers' depending on where specifically the microswimmer generates the force responsible for its propulsion. It is common to model microswimmers as a point-force dipole, as this turns out to be the leading-order contribution in the multipole expansion for the velocity field. This is represented mathematically by the stresslet tensor, defined as $S = \sigma_0 \bm{rr}$ where $S$ is the symmetric part of the first moment of the force and $\bm{r}$ is the instantaneous direction of motion. For example, the bacteria \textit{Escherichia coli} and \textit{Bacillus subtilis} move by actuating flagella positioned behind their bodies. Since the flagella 	`push' these organisms through the fluid, they are known as pushers, and their stresslet coefficient is negative: $\sigma_0<0$ \cite{saintillan2018rheology}. Other organisms, such as the marine alga \textit{Chlamydomonas reinhardtii}, are known as `pullers'	 because they propel themselves from the front ($\sigma_0>0$); for example, \textit{C. reinhardtii} swims by waving two flagella at the front of its body in a breaststroke-like fashion \cite{guasto2010oscillatory}.

Pushers and pullers affect mixing and flow in the surrounding fluid in different ways. Whereas pullers cause negligible bulk mixing and increase the viscosity compared to the liquid alone \cite{rafai2010effective}, under certain conditions, pushers cause significant mixing \cite{wu2000particle, kim2004enhanced, dunkel2013fluid} and decrease the viscosity relative to the liquid alone \cite{lopez2015turning}. If the volume fraction is high enough, pushers can cause coherent vortices to form \cite{saintillan2012emergence}, active turbulence \cite{dunkel2013fluid}, and can even lead to a superfluid-like state in which the active stresses exerted by the swimmers result in bulk fluid motion in the absence of shear \cite{lopez2015turning}. Preliminary experimental studies with artificial self-propelled particles show that such enhancement in mixing is feasible with ``pusher'' microswimmers \cite{nishiguchi2015mesoscopic} or with spherical beads isotropically coated in platinum \cite{gregory2018symmetrical}.  

In summary, there is a precedent for microswimmers (natural or synthetic) to enhance mixing in liquids. However, the ability of artificial microswimmers to enhance the transport of thermal energy has not been explored in detail. To our knowledge, the only previous studies in this vein are due to El Hasadi and Crapper \cite{el2017self,el2020self}, who formulated an analytical and numerical model of active colloid suspensions in temperature gradients, which they termed “self-propelled nanofluids,” and predicted that if the active colloids are pushers, the Nusselt number is increased (i.e., convective heat transfer rate is increased) by a factor of 3 in a square cavity exhibiting a temperature gradient perpendicular to gravity. In addition, the pusher swimmers were predicted to cause a reduction in the suspension viscosity, which is qualitatively consistent with previous rheological studies with pushers \cite{gachelin2013non, lopez2015turning}. El Hasadi and Crapper theoretically modeled “artificial pushers” that are fabricated from artificial bacterial flagella, which can be driven using an external magnetic field \cite{ghosh2009controlled, schamel2014nanopropellers}. However, the thermal transport phenomena at play at the nanoscale in active heat transfer suspensions have yet to be studied in detail. 

In this work we present molecular dynamics (MD) simulations of self-propelled nanoparticles and the effect of their motion on thermal transport through the fluid. We find that the nanoparticles’ self-propulsion leads to an overall enhancement of the fluid’s effective thermal conductivity (heat transfer rate divided by temperature gradient magnitude), as compared to the cases of Brownian (i.e., non-swimming) nanoparticles. The simulations presented here provide a theoretical foundation for this concept and lay the groundwork for future experimental studies. Since we expect that they could be useful in a broad range of heat transfer applications, we suggest the term “active heat transfer fluids” (AHTF) to describe these suspensions.

The article is arranged as follows. In Section 2 we provide a theoretical description of the problem to be solved and present the details of the MD simulation method we used. Section 3 shows the computational results and discussion. In Section 4 we present the major conclusions and discuss directions for future research, as well as  potential practical applications. 

\section{Modeling Methodology}
Understanding heat transfer enhancements due to self-propelled particles necessitates the study of nanoparticle dynamics with applied propulsive and restoring forces. Therefore, we first focus on quantifying the translational and rotational diffusivity of nanoparticles under different driving conditions using one nanoparticle. Subsequently, for thermal  conductivity measurements, we use a system with 8 nanoparticles. 

\subsection{Simulation setup and Interatomic interactions}
The nanoparticles are modeled by Lennard-Jones (LJ) atoms carved out of a perfect face-centered-cubic (FCC) crystal with lattice constant of 1.37$\sigma$. Characteristic length and energy scales used in the description of interactions between nanoparticle atoms are denoted by $\sigma$ and $\epsilon$ respectively. The crystalline atoms forming the nanoparticles are bonded together via a FENE (finitely extensible nonlinear elastic) \cite{kremer1990dynamics} potential which has the  form, 

\begin{equation}
U_{FENE}(r) = -0.5 K R_0^2 \ln \left[ 1 - \left( \frac{r}{R_0} \right)^2     \right] + 4\epsilon \left[ \left( \frac{\sigma}{r}  \right)^{12} - \left( \frac{\sigma}{r}  \right)^6   \right] +\epsilon.
\label{eqn:fene}
\end{equation}
We use $K=30\epsilon/\sigma^2$, $R_0=1.5\sigma$, while both $\sigma$ and $\epsilon$ are equal to one by definition of the unit system. The first term of Equation \ref{eqn:fene} is attractive and extends up to $R_0$; repulsive forces are represented by the 2nd term (LJ) which is cut-off at $2^{1/6}\sigma$. The FENE potential ensures that the atoms form a stable FCC crystal regardless of the temperature or strength of the LJ potential.
Each nanoparticle consists of 429 solid atoms, forming an approximately spherical shape with a radius of $\sim4\sigma$. For computational efficiency, all non-bonded interactions (liquid-liquid, nanoparticle-liquid, and atoms within the nanoparticle) are  represented by the repulsive LJ model (WCA potential) \cite{andersen1971relationship},

\begin{equation}
U_{LJ}(r) =  4\epsilon \left[ \left( \frac{\sigma}{r}  \right)^{12} - \left( \frac{\sigma}{r}  \right)^6   \right] +\epsilon,
\label{eqn:lj}
\end{equation}
which is truncated at a separation distance of $2^{1/6}\sigma$. Inclusion of attractive interactions would change the properties of the fluid as well as the interactions at the fluid-particle interface. However, in this work we look at effects of particle diffusion and propulsion on the thermal transport with respect to a generic base fluid. Therefore, we expect that the relative change in thermal conductivity will be not significantly affected by the choice of any specific fluid model.  The mass of each liquid atom is $m$ and the mass of each nanoparticle atom is $3m$. The number density of the atoms constituting the nanoparticle is over 2 times larger than that of the liquid. As a result, the nanoparticles are roughly 6 times denser than the liquid. 

All simulations are conducted using the LAMMPS MD package \cite{plimpton1995fast} using a time step size of 0.01$\sqrt{m \sigma^2/\epsilon}$. One nanoparticle simulations are carried out in cubical box of dimensions $20\sigma \times 20\sigma \times 20\sigma$ in the x, y, and z directions. For every 429 nanoparticle atoms, 6250 liquid atoms are used. Therefore, in the 8 nanoparticle simulations for evaluating effective thermal conductivity, 3,432 solid/nanoparticle atoms and 50,000 liquid atoms are present in the system; the simulation box sizes are scaled appropriately to maintain the density of one nanoparticle systems. Going forward, we mostly adhere to non-dimensional LJ units but occasionally references are made to appropriate dimensional counterparts when deemed necessary.

\subsection{Preparation of structures and equilibration \label{sec:prep_struct_equil}}
The nanoparticle is inserted into a liquid matrix with a pre-existing spherical cavity whose size is defined to avoid overlaps. The combined system is  equilibrated for 1 million time steps in the NPT ensemble (using Nos\'e-Hoover type equations of motion \cite{martyna1994constant}) at constant pressure of $5\epsilon/\sigma^3$ and temperature of $1.0\epsilon/k_B$, where $k_B$ is the Boltzmann constant. Subsequently, the system is equilibrated for 1 million time steps under the Nos\'e-Hoover constant volume and temperature (NVT) \cite{nose1984unified, hoover1985canonical} ensemble at temperature $1.0\epsilon/k_B$.  After the second stage of equilibration, to simulate the generation of propulsive force, we add forces of the same magnitude (ranging from $0.05\epsilon/\sigma$ to $0.5\epsilon/\sigma$) to each atom in the top hemisphere of the nanoparticle. The half/hemisphere of the nanoparticle to which force is applied is predefined before the self-propelling process, and the identities of these atoms are kept unchanged during the simulation. Going forward, the group of atoms with added force is referred to as the self-propelled half of the nanoparticle. The force applied to each solid atom, $\bm{F}$, is directed along the line connecting the atom with the center of mass (COM) of the nanoparticle. Consequently, the direction of the net propulsive force (\bm{$F_{np}$}) is always perpendicular to the plane separating the propelled and non-propelled halves. The magnitude of the force is constant, but its orientation varies in time because of rotational Brownian fluctuations: Specifically, the unit vector representing particle orientation performs a random walk on the unit sphere \cite{saintillan2018rheology}. 

\begin{figure}
\centering
\includegraphics[width=\linewidth]{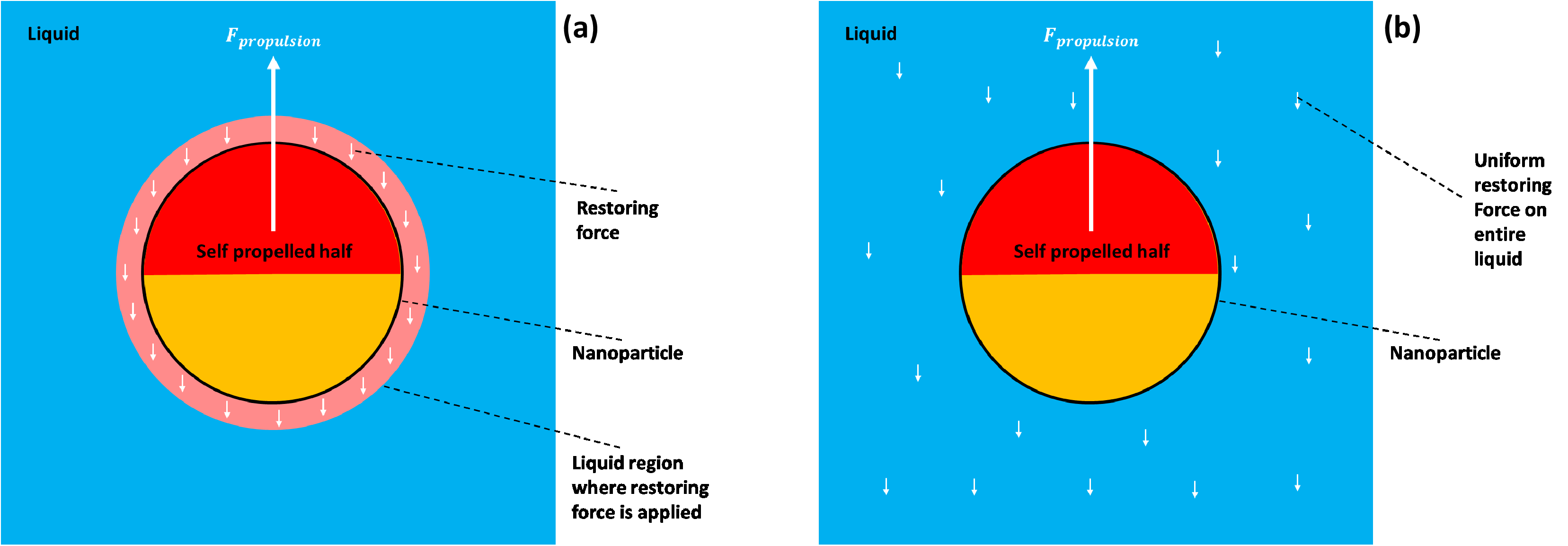}
\caption{2D Schematic of simulations. $\bm{F_{propulsion}} \equiv \bm{F_{np}}$ is the net propulsion force applied on the nanoparticle. (a) Restoring force is applied on liquid atoms which lie in a spherical shell $2\sigma$ away from the nanoparticle (b) Restoring force is applied on all liquid atoms.}
\label{fig:schematic}
\end{figure}

To imitate the effect of self-propelled particles on the fluid,  it is necessary to apply a balancing force on the liquid. In this study, we do not adhere to any specific type of propulsion mechanism, but rather consider the theoretical limits of how the balancing force could be applied so that the fluid and particle system is force-free \cite{purcell1977life, ten2015can}. We explore two limits,  (a) Local: Force is applied to liquid atoms in a region, with thickness $2\sigma$, around the nanoparticle, i.e.,  for each liquid atom,  $\bm{F_{liquid}} = - \bm{F_{np}}/N_{2\sigma}$.  $N_{2\sigma}$ is the number of atoms within $2\sigma$ of the surface. This strategy is a model of so-called ``phoretic swimmers'' that move by generating gradients in their immediate vicinity \cite{moran2017phoretic}; as shown by Golestanian et al. \cite{golestanian2007designing}, these gradients lead to body forces and slip flows that, in many practical situations, are confined to a very thin fluid layer near the particle's surface (such as an electrical double layer). (b) Global: Force is applied to all liquid atoms, i.e., $\bm{F_{liquid}} = - \bm{F_{np}}/N_{liquid}$.  $N_{liquid}$ is the total number of liquid atoms. A global application of restoring force resembles propulsion systems wherein the balancing  force is applied in a region extending beyond the immediate vicinity of the surface, for e.g., motile microorganisms \cite{lauga2020fluid} or active colloids with very thick electrical double layers (i.e., at low salt concentrations, where many active colloids move fastest) \cite{paxton2006catalytically, moran2014role}. This mechanism of application of balancing force might not be completely physically relevant but can be used as an estimate for the upper limit for our analysis. Since the restoring force is applied symmetrically about the front and back halves of the particle, the hydrodynamic signature associated with these particles is roughly that of a ``neutral'' swimmer \cite{guzman2016fission, marchetti2013hydrodynamics, elgeti2015physics}. Figure \ref{fig:schematic} shows a schematic of the single nanoparticle system and details the two methodologies  for application of the restoring force. As a control case, we also studied the system consisting of a nanoparticle without any propulsive forces (standard Brownian motion) . 

\subsection{Characterization of dynamics}
The rotational and translational motion of the nanoparticles directly determines the modification of their effective diffusivity as compared to standard Brownian motion. Prior to data collection, to reduce spurious correlations, the single nanoparticle system is further equilbrated for 22 million timesteps in the NVT ensemble. Subsequently, coordinates and velocities of the center of mass (COM) of the entire nanoparticle as well as the self-propelled half are collected for 20 million timesteps with a frequency of 5 timesteps. 

The COM velocity autocorrelation function of the nanoparticle,

\begin{equation}
\overline{\phi}_T(\tau) = \frac{\langle v_\alpha(0)v_\alpha(\tau)\rangle}{3}
\label{eqn:vel_auto}
\end{equation}
can be utilized to evaluate the effective translational diffusivity using the Green-Kubo relations \cite{green1954markoff,kubo1957statistical}, 

\begin{equation}
D_{eff} = \int_{0}^{\infty}\overline{\phi}_T(\tau) \rm{d} \tau.
\label{eqn:green_kubo}
\end{equation}
In Equation \ref{eqn:vel_auto}, $\alpha = x, y, z$ represents the three Cartesian coordinates and we are using Einstein notation (thus summation over the three indices is implied). The triangular brackets indicate averaging over multiple time origins. An estimate of relaxation time for this autocorrelation function can be obtained by fitting exponentials to the normalized functions: 

\begin{equation}
\begin{split}
\phi_T(\tau) = \frac{\overline{\phi}_T(\tau)}{\overline{\phi}_T(0)} = \exp (-\frac{\tau}{\tau_T}) ~~~~~\rm{(Brownian)} \\
\phi_T(\tau) = \frac{\overline{\phi}_T(\tau)}{\overline{\phi}_T(0)} = a\exp (-\frac{\tau}{\tau_{T1}}) + b\exp (-\frac{\tau}{\tau_{T2}}) ~~~~~\rm{(Self-propelled)}.
\label{eqn:relaxation_time_def} 
\end{split}
\end{equation}
The existence of two relaxation times for the self-propelled particles is discussed in Section \ref{sec:local_app} and shown in Figure \ref{fig:diffusivity_separate}.

During production runs, to estimate reorientation time of the nanoparticle, a vector  pointing from the COM of self-propelled half to the COM of whole nanoparticle, $\bm{r}(t)$, is calculated. The autocorrelation function for this vector is defined as, 

\begin{equation}
\overline{\phi}_R(\tau) = \frac{\langle r_\alpha(0)r_\alpha(\tau)\rangle}{3},
\label{eqn:reorient_auto}
\end{equation}
where again the index $\alpha$ indicates each of the Cartesian coordinate directions and the summation is implied. The characteristic reorientation time, $\tau_R$, is obtained by fitting an exponential to the reorientation autocorrelation function in the form, 

\begin{equation}
\begin{split}
\phi_R(\tau) = \frac{\overline{\phi}_R(\tau)}{\overline{\phi}_R(0)} = \exp (-\frac{\tau}{\tau_R}).
\end{split}
\end{equation}
\subsection{Effective Thermal Conductivity \label{sec:effective_kappa}}
To determine the effect of  self-propulsion on effective thermal conductivity ($\kappa_{eff}$), we use 8 nanoparticles in a box with 50,000 liquid atoms. After the system is  equilibrated, we employ a direct method to calculate the thermal conductivity, in which a planar heat source and sink are inserted into the domain with the overall thermostat still enabled.  The heat source and sink were placed at 1/4 of the box length from opposite edges of the box. In addition, we enforce periodic boundary conditions in all directions. Overall, a temperature gradient is established between the heat source and heat sink; a gradient equal in magnitude but opposite in sign is also established between the heat source and its nearest edge and between the heat sink and its nearest edge. Estimates of temperature gradients are obtained by averaging over the two slopes established in our systems (see Figures \ref{fig:effective_kappa} and \ref{fig:allLiq_effective_kappa}). For simplicity, while presenting results, we shift our horizontal axis such that the source is at the center and sinks are at edges of our simulation box. 

The velocities of atoms in the heat source and sink are increased and decreased, respectively, at a constant rate of   $200\epsilon/\sqrt{m \sigma^2/\epsilon}$. This rate is equal to the overall heat transfer rate through the fluid, $\dot{Q}$ . The temperature profile in the system is then calculated by determining the average kinetic energy of the atoms as a function of position. After the temperature gradient, $\rm{d}T/\rm{d}z$, reaches a steady value, the effective thermal conductivity is determined through Fourier’s Law:

\begin{equation}
\kappa_{eff} = - \frac{\dot{Q}}{2A(\rm{d}T/\rm{d}z)},
\label{eqn:keff}
\end{equation}
where $A$ represents the cross sectional area perpendicular to the temperature gradient. The factor of 2 in the denominator of Equation \ref{eqn:keff} is necessary because the total energy consumption/insertion rate, $\dot{Q}$ , flows evenly to either side of the heat source/sink. Accordingly, the net heat transfer rate from the heat source to each heat sink is $\dot{Q}/2$. Since these simulations are performed in the NVT ensemble, the net heat input/output to the system is only approximately equal to $\dot{Q}$. We use a thermostat to eliminate viscous heating of the system. Since heat is on an average generated uniformly throughout the system and the thermostat also removes this heat uniformly, effect of performing simulations in the canonical ensemble on temperature gradients is to first approximation negligible. While the absolute value of $\kappa_{eff}$ might be modified due to this correction, the relative enhancement (w.r.t Brownian) should be consistent with the results reported without the correction. 

\section{Results and Discussion}
Our modeling approach involves using MD simulations in finite simulation box sizes. Therefore, before presenting our results on diffusivity and thermal conductivity enhancements,  we briefly discuss the effect of system size on autocorrelation functions. Following this, the effect of propulsive forces on diffusivity and thermal conductivity is explored in the two limiting cases, as described in Section \ref{sec:prep_struct_equil}: Local and Global application  of balancing force on the liquid. All results are presented in non-dimensional Lennard-Jones units.

\subsection{Finite size effects}
\begin{figure}[H]
\centering
\includegraphics[width=0.5\linewidth]{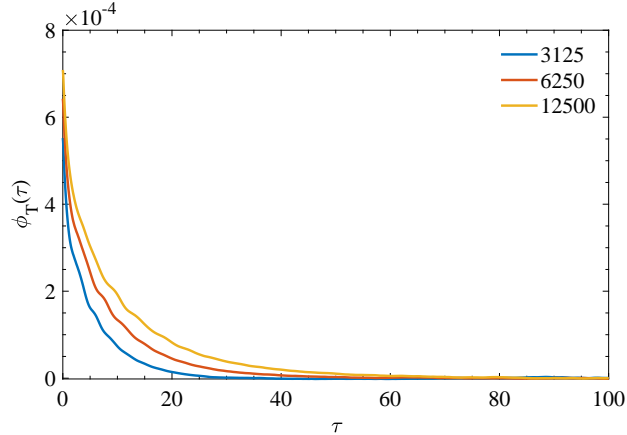}
\caption{Effect of system size on autocorrelation function of the translational velocity of COM of a nanoparticle  undergoing Brownian motion. Number of atoms in the nanoparticle is fixed at 429, while the number of atoms in the liquid is varied; the system size is scaled accordingly to preserve the liquid density.}
\label{fig:size_effect}
\end{figure}

To study the effect of system size, we consider a single Brownian nanoparticle immersed in a box of 3125, 6250, and 12500 liquid atoms. Box sizes are scaled to preserve the liquid density. As shown in Figure \ref{fig:size_effect}, the autocorrelation function of the center-of-mass (COM) velocity decays at a different rate for systems of different sizes. This size effect is well-documented and understood in the literature within a context of hydrodynamic interactions and momentum conservation \cite{asta2017transient}. We expect the size effect will be reduced if the system contains multiple nanoparticles, as hydrodynamic interactions are screened. The simulations in the remainder of the paper are conducted with 6,250 liquid atoms per solid nanoparticle.
\subsection{Local application of balancing force \label{sec:local_app}}
\subsubsection{Effective Diffusivity}
The direction of movement of self-propelled nanoparticles fluctuate stochastically in time due to rotational Brownian motion, which is superimposed on translational self-propelled motion. Thus, in the absence of an external gradient or magnetic field, the trajectory of artificial microswimmers generally resembles that of an `enhanced random walk', i.e., Brownian motion with enhanced effective diffusivity. This model is often referred to as the Active Brownian Particle (ABP) model \cite{romanczuk2012active}. The effective diffusivity of unguided active colloids depends quadratically on the self-propelled speed \cite{takatori2016forces, saintillan2018rheology}. To verify that self-propulsion is accurately represented in  MD simulations,  we first aim to quantify the effective diffusivity of the self-propelled particles.

\begin{figure}[H]
\centering
\includegraphics[width=\linewidth]{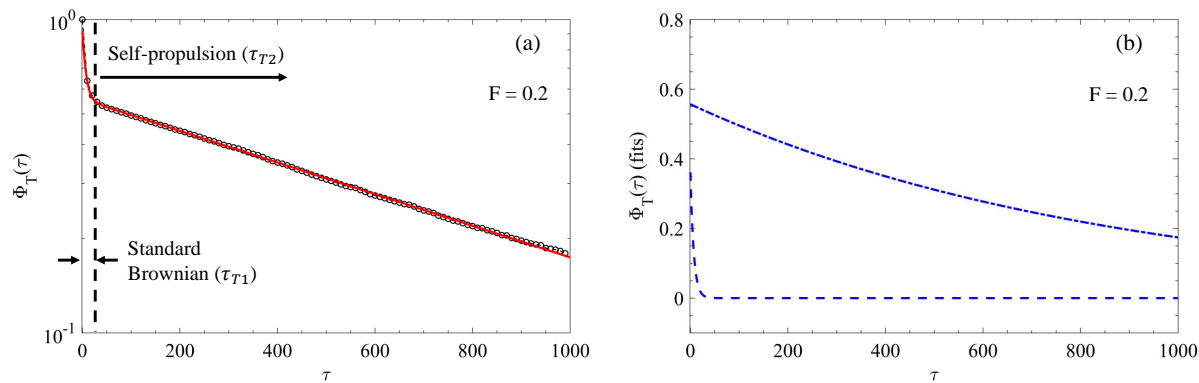}
\caption{Normalized autocorrelation function, $\phi_T(\tau)$ of COM velocity of nanoparticle when a propulsive force of 0.2 is applied to each atom in the self-propelled half. (a) Red solid line is a fit, using two exponentials, to the actual autocorrelation function denoted by black dashed line. (b) The two exponential fitting functions are plotted explicitly to show the effect of the two relaxation times -- $\tau_{T1}$ and $\tau_{T2}$.}
\label{fig:F_0.2_fitting}
\end{figure}

\begin{figure}[H]
	\centering
	\includegraphics[width=\linewidth]{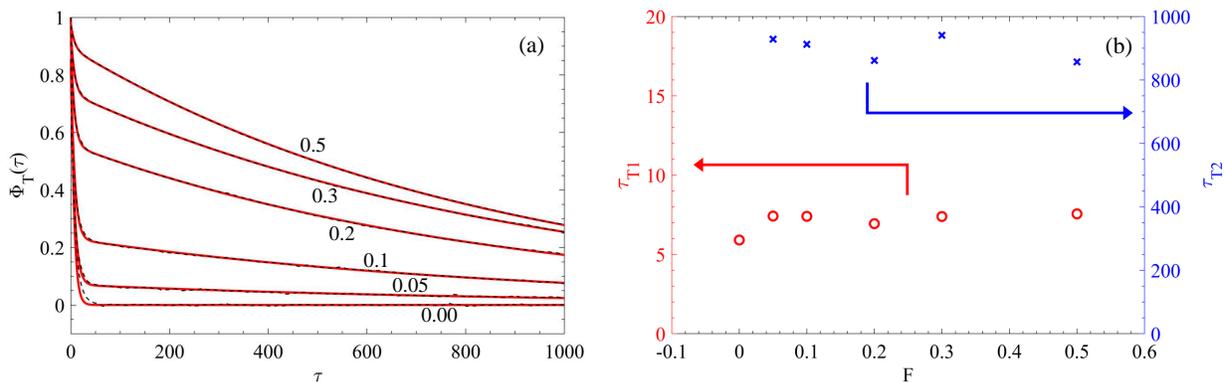}
	\caption{(a) Autocorrelation function  of COM velocity of nanoparticle as the propulsion force (green labels) is varied from 0.00 (Brownian) to 0.5. (b) Two relaxation times corresponding to the autocorrelation functions: Brownian ($\tau_{T1} $) on  left axis and Self-propulsion ($\tau_{T2} $) on  right axis. For the Brownian case, only one exponential is necessary for fitting with relaxation time $\tau_{T1} $. }
	\label{fig:relax_time}
\end{figure}

\begin{figure}[H]
\centering
\includegraphics[width=\linewidth]{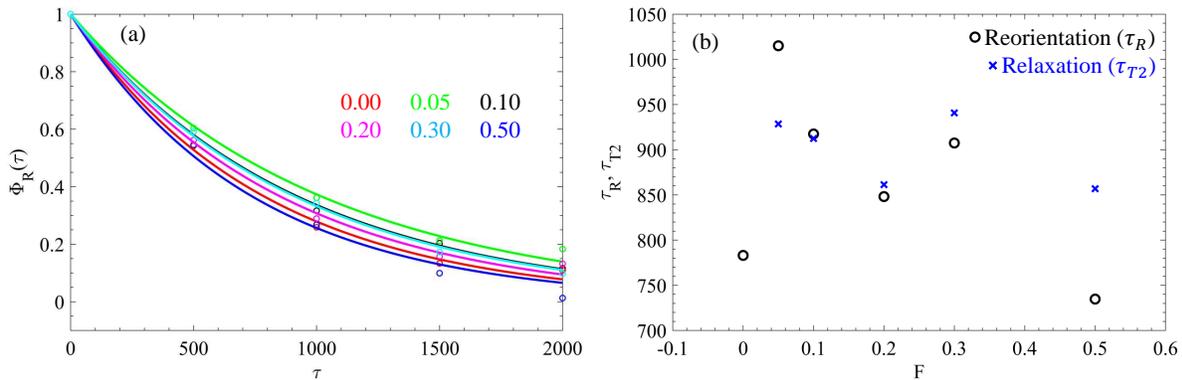}
\caption{(a) Autocorrelation function  of reorientation vector of nanoparticle as the propulsion force is varied from 0.00 (Brownian) to 0.5. (b) Relaxation times, $\tau_R$, corresponding to the autocorrelation functions. $\tau_{T2}$ is the translational relaxation time obtained from Figure \ref{fig:relax_time}(b)}
\label{fig:reorient_time}
\end{figure}

\begin{figure}[H]
	\centering
	\includegraphics[width=0.5
	\linewidth]{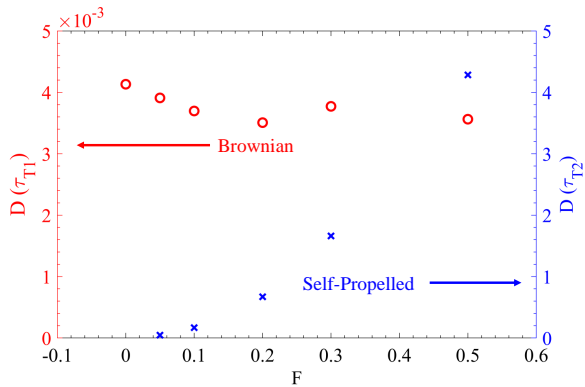}
	\caption{Diffusivity evaluated from the two exponentials used to fit the autocorrelation functions. Left axis is used to denote the diffusivity calculated from the exponential with relaxation time $\tau_{T1}$, i.e., the Brownian contribution to effective diffusivity. The right axis corresponds to the diffusivity contribution from self-propulsion (exponential with relaxation time $\tau_{T2}$). }
	\label{fig:diffusivity_separate}
\end{figure}

\begin{figure}[H]
	\centering
	\includegraphics[width=0.5\linewidth]{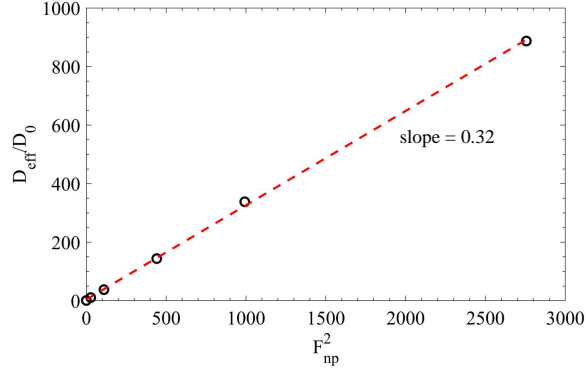}
	\caption{Ratio of self-propelled nanoparticle’s effective diffusion coefficient, $D_{eff}$, to that of the Brownian nanoparticles, $D_0$, as a function of propulsive force squared. Black symbols show simulation results, while the red dashed line indicates a linear polynomial fit. This figure confirms the quadratic relationship between the effective `swim' diffusivity and the self-propelled speed of the particles (which is linearly proportional to the force) }
	\label{fig:diffusivity}
\end{figure}

To investigate the diffusion of self-propelled particles, we analyze the autocorrelation function of its COM velocity. Figure \ref{fig:F_0.2_fitting} shows the fitting process for the case when a propulsive force, $F=0.2$ ($F=0.2\epsilon/\sigma$ in real units) is applied to each atom in the self-propelled half. Existence of two `decay' regimes for the autocorrelation function is clearly portrayed in Figure \ref{fig:F_0.2_fitting}(a). While the black dotted line represents the actual autocorrelation function, the red solid line denotes a fit using two exponentials. The functional form of the two exponential fitting functions is shown in Equation \ref{eqn:relaxation_time_def} and is explicitly plotted in Figure \ref{fig:F_0.2_fitting}(b). The two relaxation times, $\tau_{T1}$ and $\tau_{T2}$, correspond to the Brownian and self-propelled motion of the nanoparticle respectively. Autocorrelation functions and relaxation times for other values of applied forces, including the Brownian case, are shown in Figure \ref{fig:relax_time}.  Relaxation times of the self-propelled nanoparticles are significantly longer (by roughly two orders of magnitude) than purely Brownian nanoparticles. Furthermore, for self-propelled nanoparticles the relaxation times are independent of the magnitude of propulsion forces. 
To explore the independence of relaxation times w.r.t.  propulsive forces, we evaluate the orientation vector normalized autocorrelation function, $\phi_R(\tau)$, and associated reorientation time, as shown in Figure \ref{fig:reorient_time}. Notably, the self-propelled and Brownian particles are characterized by the same reorientation time; therefore, they exhibit the same behavior in terms of reorientation. This is expected because the directions of net applied forces are along the COM of the nanoparticle and produce zero torque; hence, rotation is unaffected. The exponential fits used in Figure \ref{fig:reorient_time}(a) have a single free parameter ($\tau_R$), as described in Equation \ref{eqn:relaxation_time_def}.  Furthermore, the reorientation time of self-propelled particles is approximately same as the relaxation time; $\tau_{T2}$, of COM velocity autocorrelation function (see Figure \ref{fig:relax_time}). Thus, when the particles propel themselves, the relaxation time for translational motion is dominated by orientation decorrelation, as the translational diffusion is dominated by the propulsion velocity, not Brownian velocity. 
Although the discussion on relaxation times is crucial for understanding the dynamics of the system, evaluation of diffusion coefficients is equally important. In Figure \ref{fig:diffusivity_separate}, we separately plot the diffusivities calculated (using Equation \ref{eqn:green_kubo}) from the two exponential fits to the COM velocity autocorrelation functions shown in Figure \ref{fig:relax_time}(a). The Brownian diffusivity, calculated from exponential with relaxation time $\tau_{T1}$, is indeed independent of the propulsive force as envisioned earlier. A theoretical estimate of this diffusivity can be obtained using the Stokes-Einstein relationship, $D_0=k_BT/6 \pi \eta R$ -- where $\eta$ and $R$ are the viscosity of fluid and particle radius respectively.  In LJ units, the repulsive LJ potential (WCA) \cite{andersen1971relationship} yields a dynamic viscosity of $\sim$2.24 \cite{ohtori2018stokes} at the considered density. Therefore, the estimate of $D_0$ becomes $\sim 6 \times 10^{-3}$ and is commensurate with observed Brownian diffusivities reported in Figure \ref{fig:diffusivity_separate}. On the right axis of Figure \ref{fig:diffusivity_separate}, we plot the contribution to diffusivity due to self-propulsion, $D_{swim}$, which shows a non-linear behavior with applied force. To explore this, we next investigate the variation of $D_{eff}$ from our simulations and a simplified theoretical analysis. 
In Figure \ref{fig:diffusivity} we show the ratio of diffusion coefficients of self-propelled nanoparticles to that of Brownian nanoparticles,  $D_{eff}/D_0$. $F_{np}$ is the net force on the nanoparticle that is calculated by summing the forces on each atom in the self propelled half while considering the specific component cancellations due to spherical symmetry of the nanoparticle. It should be noted that the diffusivity values reported in this Section are calculated by explicitly integrating the autocorrelation functions and not the exponential fits shown in Figure \ref{fig:diffusivity}. The large enhancement in effective diffusivity is a common observation in the self-propelled particles literature and the quadratic dependence of the effective diffusivity on linear speed is also well-established \cite{saintillan2018rheology, howse2007self}. In a recent simulation study, El Hasadi and Crapper found that effective diffusivity is enhanced by several orders of magnitude compared to the Brownian value as speed is increased, and depends non-monotonically on size (diffusivity decreases with increasing size for particle sizes below a certain speed-dependent threshold size, and increases with increasing size above this threshold) \cite{el2020self}. Experimentally, the ratio of swim to Brownian diffusivity has been observed to be greater than 1000 \cite{takatori2016forces} and is comparable to  those reported in Figure \ref{fig:diffusivity}.
To understand Figure \ref{fig:diffusivity} better, we perform a simple analysis.  Effective diffusivity of self-propelled nanoparticles can be related to the self-propelled speed $V_0$ \cite{saintillan2018rheology, el2020self, takatori2016forces} by,

\begin{equation}
D_{eff}= D_0+\frac{V^2_0}{6D_R}
\label{eqn:diff_expr}
\end{equation}
where  $D_0$ is the Brownian diffusivity, $D_R=k_B T/ 8 \pi \eta R^3$ is the rotational diffusivity of the (spherical) particle, $k_BT$ is the thermal energy, $\eta$ is the dynamic viscosity of the fluid, and R is the particle radius. The force applied on the nanoparticle, $F_{np}$ should be proportional to the velocity, $V_0$. One possible way of relating these two is using the expression for Stokes' drag  ($F_{np}=6 \pi \eta R V_0$), that is valid under the assumptions: Reynolds Number (Re) is small and disturbance on fluid in the vicinity of the nanoparticle is similar to standard Brownian motion. Using this relation, Equation \ref{eqn:diff_expr} can be written as, 

\begin{equation}
D_{eff} = D_0 + \frac{RF^2_{np}}{27 \pi \eta k_B T}.
\label{eqn:diff_expr1}
\end{equation}
According to the Stokes-Einstein relationship, $D_0=k_BT/6 \pi \eta R$. Therefore the estimate of $D_{eff}/D_0$ becomes, 

\begin{equation}
\frac{D_{eff}}{D_0}  =1+ \frac{2}{9} \left( \frac{R}{k_B T} \right)^2 F^2_{np} \approx 1 + 3.5 F^2_{np},
\label{eqn:diff_expr2}
\end{equation}
where we have used the fact that, in the MD reduced unit system, the thermal energy $k_BT = 1$ and $R\approx 4$.
Although the pre-factor in above Equation differs from the slope in Figure \ref{fig:diffusivity}, the functional dependence on $F_{np}$ is the same. This indicates the consistency of observations based on MD simulations and theoretical analysis of propelled and standard Brownian motion. Discrepancy in the pre-factor could be due to the assumptions made in the analysis, specifically Stokes drag -- due to local application of balancing forces, the hydrodynamics near and far away from the nanoparticle could be different from what is assumed in deriving Stokes drag. Another point of deviation from Stokes' analysis is due to finite size of simulation box with periodic BCs, which indicates that fluid far away from nanoparticle is not stagnant. Additionally, the existence of this deviation could be more fundamental, for e.g., presence of active turbulence \cite{bechinger2016active, thampi2016active} due to high propulsion forces in our system. Further investigations in these directions are necessary to ascertain the exact cause of this deviation.

\subsubsection{Effective Thermal Conductivity}
\begin{figure}[H]
	\centering
	\includegraphics[width=0.7\linewidth]{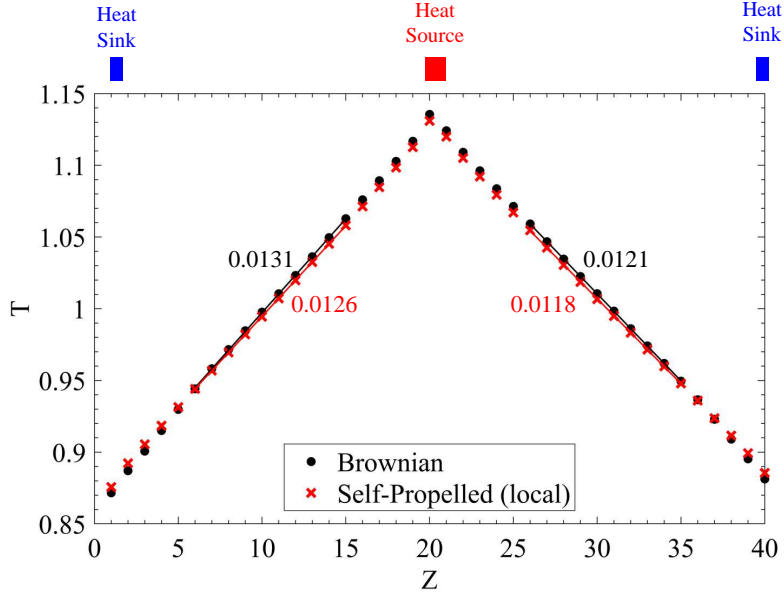}
	\caption{Temperature profile in 8 nanoparticle system undergoing Brownian and self-propelled motion when the balancing force is applied locally. Straight lines, both black and red, represent the fits used to obtain the temperature gradient estimates. The numbers in red (black) represents the magnitude of temperature gradient for the self-propelled (Brownian) systems. }
	\label{fig:effective_kappa}
\end{figure}

Effective diffusion constants for systems undergoing self-propelled motion can be orders of magnitude larger than that of standard Brownian motion, as shown in the previous Section. This suggests that enhanced (propelled) Brownian motion might increase effective thermal conductivity. To test this hypothesis we evaluate thermal conductivity of the system using the methodology described in Section \ref{sec:effective_kappa}. In Figure \ref{fig:effective_kappa}, the steady state temperature profiles for planar heat source-sink simulations of the 8 nanoparticle systems is shown. Average slopes for Brownian and self-propelled cases are 0.0126 and 0.0122 respectively. Using this data in Equation \ref{eqn:keff}, we observe a 3.1\% increase in effective thermal conductivity for an applied force of $F=0.5$ as compared to systems wherein nanoparticles undergo pure Brownian motion. 

\subsection{Global application of balancing force}
Local application, within $2\sigma$, of balancing force on the liquid was one of the limiting ways the restoring force could be applied. The other limit is when all the liquid atoms in the simulation box experience the restoring force. We discuss this scenario here. In general, all conclusions drawn in Section \ref{sec:local_app} are applicable here. 
\subsubsection{Effective diffusivity}
\begin{figure}[H]
	\centering
	\includegraphics[width=\linewidth]{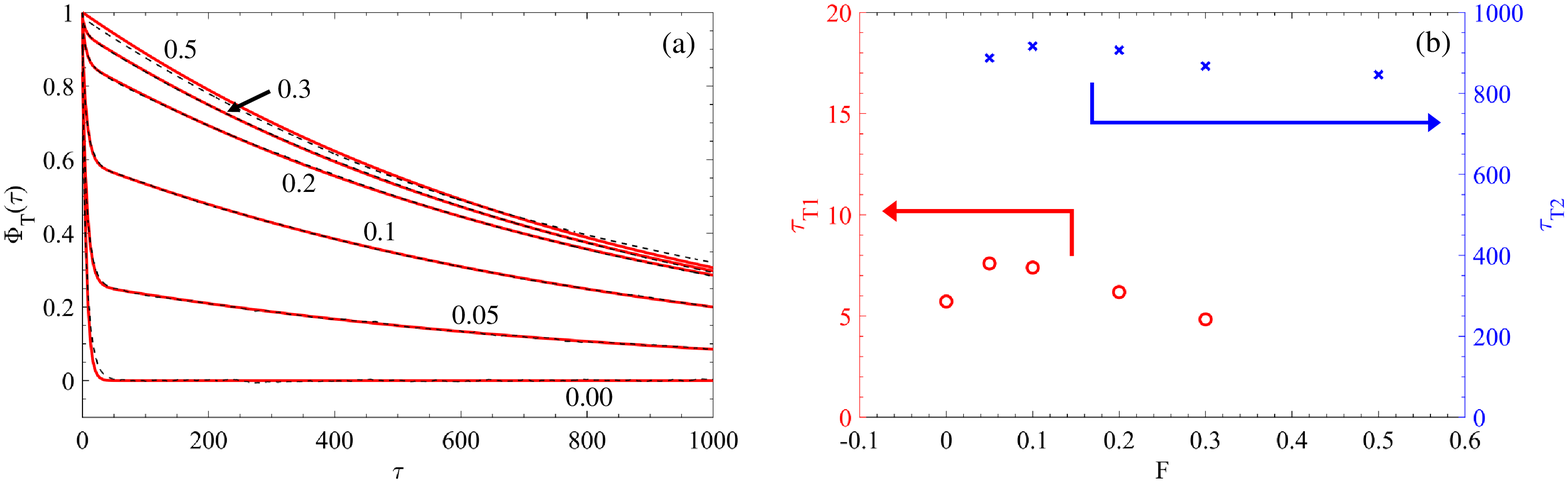}
	\caption{(a) Autocorrelation function  of COM velocity of nanoparticle as the propulsion force (green labels) is varied from 0.00 (Brownian) to 0.5. (b) Two relaxation times corresponding to the autocorrelation functions: Brownian ($\tau_{T1} $) on  left axis and Self-propulsion ($\tau_{T2} $) on  right axis. For the Brownian case, only one exponential is necessary for fitting, $\tau_{T1} $. When $F=0.5$, only one exponential, $\tau_{T2}$, is used for fitting as the inital Brownian regime is almost non-existent due to the sampling frequency considered in our data.  }
	\label{fig:allLiq_relax_time}
\end{figure}

\begin{figure}[H]
	\centering
	\includegraphics[width=\linewidth]{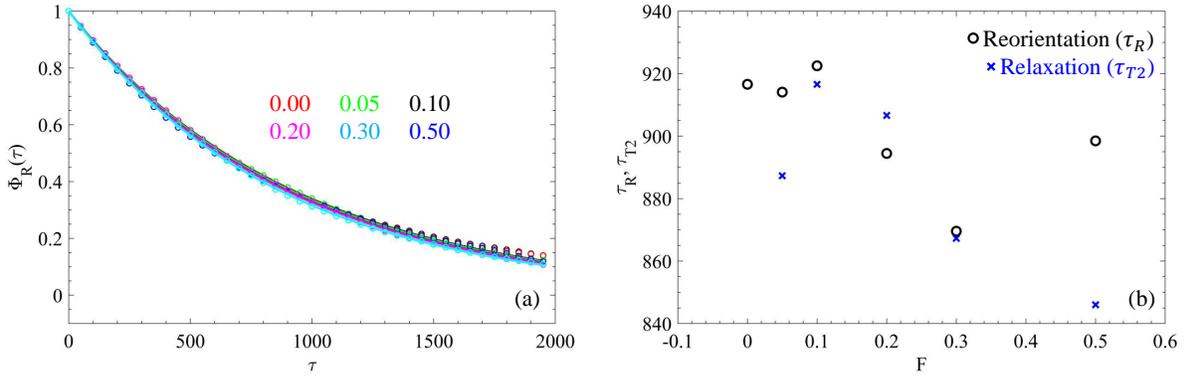}
	\caption{(a) Autocorrelation function  of reorientation vector of nanoparticle as the propulsion force is varied from 0.00 (Brownian) to 0.5. (b) Relaxation times, $\tau_R$, corresponding to the autocorrelation functions. $\tau_{T2}$ is the translational relaxation time obtained from Figure \ref{fig:allLiq_relax_time}(b).}
	\label{fig:allLiq_reorient_time}
\end{figure}

Figure \ref{fig:allLiq_relax_time} shows the autocorrelation functions are the corresponding relaxation times. With the exception of $F=0$ (Brownian) and $F=0.5$, two exponentials are used for obtaining the fits. When, $F=0.5$ a single exponential is used for fitting as the inital Brownian regime is almost non-existent due to our data sampling frequency. The relaxation times obtained here are comparable to the those presented in Section \ref{sec:local_app}; thus, the dynamics of nanoparticle does not get altered by how the balancing force is applied. This is also reinforced by the similarity in reorientation times as shown in Figure \ref{fig:allLiq_reorient_time}. However, the magnitudes of effective diffusivity warrants some discussion.

\begin{figure}[H]
	\centering
	\includegraphics[width=0.7\linewidth]{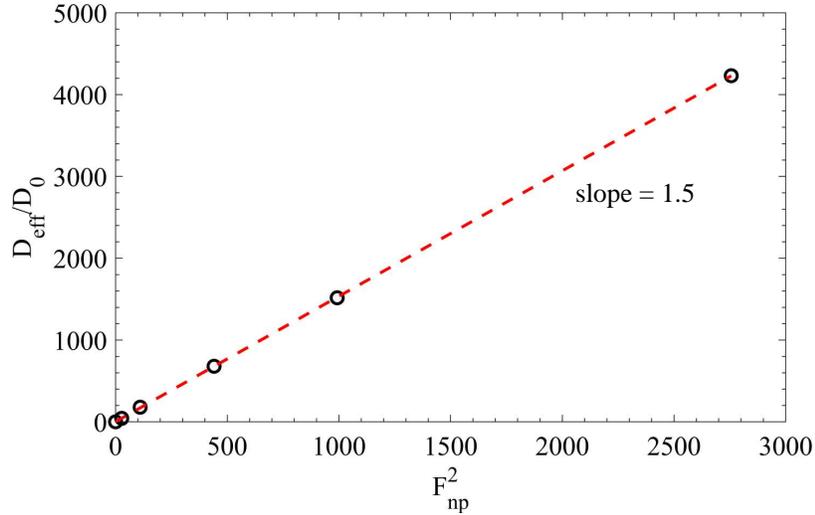}
	\caption{Ratio of self-propelled nanoparticle’s effective diffusion coefficient, $D_{eff}$, to that of the Brownian nanoparticles, $D_0$, as a function of propulsive force squared. Black symbols show simulation results, while the red dashed line indicates a linear polynomial fit. The slope is $\sim$4 times larger than that obtained from the earlier set of simulations wherein balancing forces are applied locally. }
	\label{fig:allLiq_diffusivity}
\end{figure}

In Figure \ref{fig:allLiq_diffusivity}, the dependence of $D_{eff}/D_0$ with $F^2_{np}$ is shown. While the functional dependence is consistent with that of Figure \ref{fig:diffusivity} and theoretical considerations (Equations \ref{eqn:diff_expr} - \ref{eqn:diff_expr2}), the slope is larger. In fact, with the global application of balancing force, the slope is closer to that obtained from theory (using Stokes drag). The value of diffusivity is roughly 4 times higher as compared to the systems when balancing force is applied locally. This can be attributed to the fact that, in a global balance of forces, the COM velocity of the nanoparticle is `more' correlated: thus, the autocorrelation function shifts upward. Relaxation times are not modified significantly because they are solely dependent on reorientation times, and reorientation mechanisms are unaltered with the global application of force.

\subsubsection{Effective Thermal Conductivity}
\begin{figure}[H]
	\centering
	\includegraphics[width=0.7\linewidth]{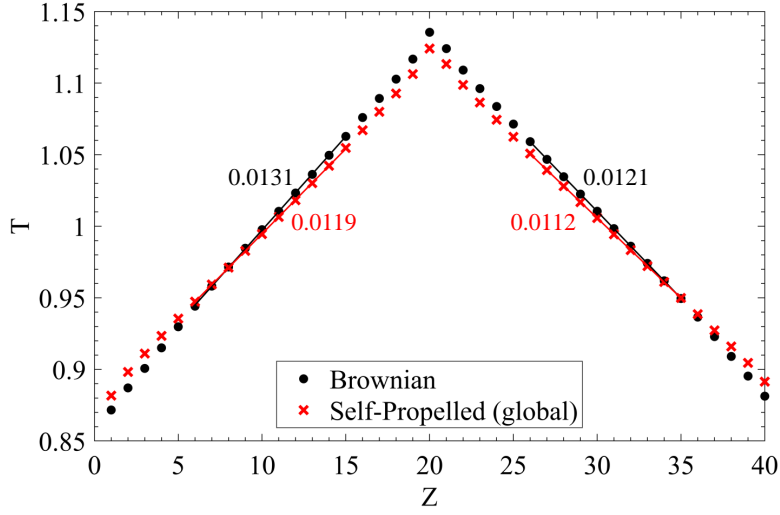}
	\caption{Temperature profile in 8 nanoparticle system undergoing Brownian and self-propelled motion where the balancing force on liquid is applied globally. The numbers in red (black) represents the magnitude of temperature gradient for the self-propelled (Brownian) systems. Straight lines, both black and red, represent the fits used to obtain the temperature gradient estimates.}
	\label{fig:allLiq_effective_kappa}
\end{figure}

In lines with our earlier discussion in Section \ref{sec:local_app}, we now focus on the enhancement in effective thermal conductivity when the balancing force is applied globally. In Figure \ref{fig:allLiq_effective_kappa} we show the steady state temperature profiles for planar heat source-sink simulations of 8 nanoparticle model nanofluids . In the presence of self-propulsion, the average temperature gradient magnitude is 0.0115. The increase in $\kappa_{eff}$ is 9.4\% as compared to the Brownian case. Since we present two limits of the how the balancing force could be applied, in a real system the increase in effective thermal conductivity would be approximately in the range: 3.7\% to 9.4\%.  However, this enhancement will scale significantly with the increase in size, speed, and volume fraction of nanoparticles. 

\section{Summary and Conclusions }
We have presented results of MD simulations of self-propelled nanoparticles and studied their effect on the temperature distribution within the fluid. Our propulsion mechanism resembles closely a ``neutral'' swimmer.  Explicitly characterizing flow fields produced by the swimming nanoparticles, especially spherical ones, is an active area of research \cite{chiang2014localized, campbell2019experimental}. To model ``pushers'' and ``pullers''  we speculate that balancing force could be applied solely to the liquid atoms adjacent to the self-propelled half \cite{saintillan2012emergence}.

First, we demonstrated that translational and rotational dynamics follow behavior expected from the theoretical analysis of standard and self-propelled Brownian nanoparticles. In the case of self-propelled particles, the relaxation time for the translational diffusion coefficient is controlled by the reorientation relaxation time, $\tau_R$. 

In the case of self-propelled nanoparticles, we observe a discriminable increase of effective thermal conductivity of the fluid. While these increases of up to several percent might be not spectacular, they are significant considering these increases are multiplicative with increases in thermal conductivity due to addition of conventional nanoparticles.  MD simulations and theoretical analysis have shown that the enhancement provided by these conventional nanoparticles derives solely from the increased conduction heat transfer \cite{maxwell1873treatise, buongiorno2009benchmark} and standard Brownian motion contributes minimally \cite {evans2006role}.  Furthermore, the nanoparticles simulated in this work have a diameter equivalent in physical units to a just few nanometers. By contrast, active colloids typically range from 100 nm to 10 $\mu$m in size. Thus, the particles simulated here are orders of magnitude smaller than typical active colloids. Heat transfer enhancement is expected to scale with size, self-propelled speed, and volume fraction of the active colloids. Exploration and experimental validation of these scaling behaviors will be the subject of future studies along with measurements of the effect of self-propulsion and temperature changes on the suspension viscosity. Our results indicate that self-propelled particles have the potential to enhance thermal conductivity and heat transfer by an amount that can enable new applications in thermal management.

\section{Acknowledgement}
This work was supported by NSF EAGER award CBET-2039262 and used the Extreme Science and Engineering Discovery Environment (XSEDE), which is supported by National Science Foundation grant number ACI-1548562.

\section*{Availability of Data}
The data that support the findings of this study are available from the corresponding author upon reasonable request.

\bibliography{biblio}

\begin{thebibliography}{55}%
\makeatletter
\providecommand \@ifxundefined [1]{%
 \@ifx{#1\undefined}
}%
\providecommand \@ifnum [1]{%
 \ifnum #1\expandafter \@firstoftwo
 \else \expandafter \@secondoftwo
 \fi
}%
\providecommand \@ifx [1]{%
 \ifx #1\expandafter \@firstoftwo
 \else \expandafter \@secondoftwo
 \fi
}%
\providecommand \natexlab [1]{#1}%
\providecommand \enquote  [1]{``#1''}%
\providecommand \bibnamefont  [1]{#1}%
\providecommand \bibfnamefont [1]{#1}%
\providecommand \citenamefont [1]{#1}%
\providecommand \href@noop [0]{\@secondoftwo}%
\providecommand \href [0]{\begingroup \@sanitize@url \@href}%
\providecommand \@href[1]{\@@startlink{#1}\@@href}%
\providecommand \@@href[1]{\endgroup#1\@@endlink}%
\providecommand \@sanitize@url [0]{\catcode `\\12\catcode `\$12\catcode
  `\&12\catcode `\#12\catcode `\^12\catcode `\_12\catcode `\%12\relax}%
\providecommand \@@startlink[1]{}%
\providecommand \@@endlink[0]{}%
\providecommand \url  [0]{\begingroup\@sanitize@url \@url }%
\providecommand \@url [1]{\endgroup\@href {#1}{\urlprefix }}%
\providecommand \urlprefix  [0]{URL }%
\providecommand \Eprint [0]{\href }%
\providecommand \doibase [0]{http://dx.doi.org/}%
\providecommand \selectlanguage [0]{\@gobble}%
\providecommand \bibinfo  [0]{\@secondoftwo}%
\providecommand \bibfield  [0]{\@secondoftwo}%
\providecommand \translation [1]{[#1]}%
\providecommand \BibitemOpen [0]{}%
\providecommand \bibitemStop [0]{}%
\providecommand \bibitemNoStop [0]{.\EOS\space}%
\providecommand \EOS [0]{\spacefactor3000\relax}%
\providecommand \BibitemShut  [1]{\csname bibitem#1\endcsname}%
\let\auto@bib@innerbib\@empty
\bibitem [{\citenamefont {Choi}\ and\ \citenamefont
  {Eastman}(1995)}]{choi1995enhancing}%
  \BibitemOpen
  \bibfield  {author} {\bibinfo {author} {\bibfnamefont {S.~U.}\ \bibnamefont
  {Choi}}\ and\ \bibinfo {author} {\bibfnamefont {J.~A.}\ \bibnamefont
  {Eastman}},\ }\href@noop {} {\enquote {\bibinfo {title} {Enhancing thermal
  conductivity of fluids with nanoparticles},}\ }\bibinfo {type} {Tech. Rep.}\
  (\bibinfo  {institution} {Argonne National Lab., IL (United States)},\
  \bibinfo {year} {1995})\BibitemShut {NoStop}%
\bibitem [{\citenamefont {Maxwell}(1873)}]{maxwell1873treatise}%
  \BibitemOpen
  \bibfield  {author} {\bibinfo {author} {\bibfnamefont {J.~C.}\ \bibnamefont
  {Maxwell}},\ }\href@noop {} {\emph {\bibinfo {title} {A treatise on
  electricity and magnetism}}},\ Vol.~\bibinfo {volume} {1}\ (\bibinfo
  {publisher} {Oxford: Clarendon Press},\ \bibinfo {year} {1873})\BibitemShut
  {NoStop}%
\bibitem [{\citenamefont {Prasher}, \citenamefont {Bhattacharya},\ and\
  \citenamefont {Phelan}(2005)}]{prasher2005thermal}%
  \BibitemOpen
  \bibfield  {author} {\bibinfo {author} {\bibfnamefont {R.}~\bibnamefont
  {Prasher}}, \bibinfo {author} {\bibfnamefont {P.}~\bibnamefont
  {Bhattacharya}}, \ and\ \bibinfo {author} {\bibfnamefont {P.~E.}\
  \bibnamefont {Phelan}},\ }\bibfield  {title} {\enquote {\bibinfo {title}
  {Thermal conductivity of nanoscale colloidal solutions (nanofluids)},}\
  }\href@noop {} {\bibfield  {journal} {\bibinfo  {journal} {Physical review
  letters}\ }\textbf {\bibinfo {volume} {94}},\ \bibinfo {pages} {025901}
  (\bibinfo {year} {2005})}\BibitemShut {NoStop}%
\bibitem [{\citenamefont {Evans}, \citenamefont {Fish},\ and\ \citenamefont
  {Keblinski}(2006)}]{evans2006role}%
  \BibitemOpen
  \bibfield  {author} {\bibinfo {author} {\bibfnamefont {W.}~\bibnamefont
  {Evans}}, \bibinfo {author} {\bibfnamefont {J.}~\bibnamefont {Fish}}, \ and\
  \bibinfo {author} {\bibfnamefont {P.}~\bibnamefont {Keblinski}},\ }\bibfield
  {title} {\enquote {\bibinfo {title} {Role of brownian motion hydrodynamics on
  nanofluid thermal conductivity},}\ }\href@noop {} {\bibfield  {journal}
  {\bibinfo  {journal} {Applied Physics Letters}\ }\textbf {\bibinfo {volume}
  {88}},\ \bibinfo {pages} {093116} (\bibinfo {year} {2006})}\BibitemShut
  {NoStop}%
\bibitem [{\citenamefont {Babaei}, \citenamefont {Keblinski},\ and\
  \citenamefont {Khodadadi}(2013)}]{babaei2013proof}%
  \BibitemOpen
  \bibfield  {author} {\bibinfo {author} {\bibfnamefont {H.}~\bibnamefont
  {Babaei}}, \bibinfo {author} {\bibfnamefont {P.}~\bibnamefont {Keblinski}}, \
  and\ \bibinfo {author} {\bibfnamefont {J.}~\bibnamefont {Khodadadi}},\
  }\bibfield  {title} {\enquote {\bibinfo {title} {A proof for insignificant
  effect of brownian motion-induced micro-convection on thermal conductivity of
  nanofluids by utilizing molecular dynamics simulations},}\ }\href@noop {}
  {\bibfield  {journal} {\bibinfo  {journal} {Journal of Applied Physics}\
  }\textbf {\bibinfo {volume} {113}},\ \bibinfo {pages} {084302} (\bibinfo
  {year} {2013})}\BibitemShut {NoStop}%
\bibitem [{\citenamefont {Eapen}\ \emph {et~al.}(2007)\citenamefont {Eapen},
  \citenamefont {Williams}, \citenamefont {Buongiorno}, \citenamefont {Hu},
  \citenamefont {Yip}, \citenamefont {Rusconi},\ and\ \citenamefont
  {Piazza}}]{eapen2007mean}%
  \BibitemOpen
  \bibfield  {author} {\bibinfo {author} {\bibfnamefont {J.}~\bibnamefont
  {Eapen}}, \bibinfo {author} {\bibfnamefont {W.~C.}\ \bibnamefont {Williams}},
  \bibinfo {author} {\bibfnamefont {J.}~\bibnamefont {Buongiorno}}, \bibinfo
  {author} {\bibfnamefont {L.-w.}\ \bibnamefont {Hu}}, \bibinfo {author}
  {\bibfnamefont {S.}~\bibnamefont {Yip}}, \bibinfo {author} {\bibfnamefont
  {R.}~\bibnamefont {Rusconi}}, \ and\ \bibinfo {author} {\bibfnamefont
  {R.}~\bibnamefont {Piazza}},\ }\bibfield  {title} {\enquote {\bibinfo {title}
  {Mean-field versus microconvection effects in nanofluid thermal
  conduction},}\ }\href@noop {} {\bibfield  {journal} {\bibinfo  {journal}
  {Physical review letters}\ }\textbf {\bibinfo {volume} {99}},\ \bibinfo
  {pages} {095901} (\bibinfo {year} {2007})}\BibitemShut {NoStop}%
\bibitem [{\citenamefont {Putnam}\ \emph {et~al.}(2006)\citenamefont {Putnam},
  \citenamefont {Cahill}, \citenamefont {Braun}, \citenamefont {Ge},\ and\
  \citenamefont {Shimmin}}]{putnam2006thermal}%
  \BibitemOpen
  \bibfield  {author} {\bibinfo {author} {\bibfnamefont {S.~A.}\ \bibnamefont
  {Putnam}}, \bibinfo {author} {\bibfnamefont {D.~G.}\ \bibnamefont {Cahill}},
  \bibinfo {author} {\bibfnamefont {P.~V.}\ \bibnamefont {Braun}}, \bibinfo
  {author} {\bibfnamefont {Z.}~\bibnamefont {Ge}}, \ and\ \bibinfo {author}
  {\bibfnamefont {R.~G.}\ \bibnamefont {Shimmin}},\ }\bibfield  {title}
  {\enquote {\bibinfo {title} {Thermal conductivity of nanoparticle
  suspensions},}\ }\href@noop {} {\bibfield  {journal} {\bibinfo  {journal}
  {Journal of Applied Physics}\ }\textbf {\bibinfo {volume} {99}},\ \bibinfo
  {pages} {084308} (\bibinfo {year} {2006})}\BibitemShut {NoStop}%
\bibitem [{\citenamefont {Nan}\ \emph {et~al.}(1997)\citenamefont {Nan},
  \citenamefont {Birringer}, \citenamefont {Clarke},\ and\ \citenamefont
  {Gleiter}}]{nan1997effective}%
  \BibitemOpen
  \bibfield  {author} {\bibinfo {author} {\bibfnamefont {C.-W.}\ \bibnamefont
  {Nan}}, \bibinfo {author} {\bibfnamefont {R.}~\bibnamefont {Birringer}},
  \bibinfo {author} {\bibfnamefont {D.~R.}\ \bibnamefont {Clarke}}, \ and\
  \bibinfo {author} {\bibfnamefont {H.}~\bibnamefont {Gleiter}},\ }\bibfield
  {title} {\enquote {\bibinfo {title} {Effective thermal conductivity of
  particulate composites with interfacial thermal resistance},}\ }\href@noop {}
  {\bibfield  {journal} {\bibinfo  {journal} {Journal of Applied Physics}\
  }\textbf {\bibinfo {volume} {81}},\ \bibinfo {pages} {6692--6699} (\bibinfo
  {year} {1997})}\BibitemShut {NoStop}%
\bibitem [{\citenamefont {Buongiorno}\ \emph {et~al.}(2009)\citenamefont
  {Buongiorno}, \citenamefont {Venerus}, \citenamefont {Prabhat}, \citenamefont
  {McKrell}, \citenamefont {Townsend}, \citenamefont {Christianson},
  \citenamefont {Tolmachev}, \citenamefont {Keblinski}, \citenamefont {Hu},
  \citenamefont {Alvarado} \emph {et~al.}}]{buongiorno2009benchmark}%
  \BibitemOpen
  \bibfield  {author} {\bibinfo {author} {\bibfnamefont {J.}~\bibnamefont
  {Buongiorno}}, \bibinfo {author} {\bibfnamefont {D.~C.}\ \bibnamefont
  {Venerus}}, \bibinfo {author} {\bibfnamefont {N.}~\bibnamefont {Prabhat}},
  \bibinfo {author} {\bibfnamefont {T.}~\bibnamefont {McKrell}}, \bibinfo
  {author} {\bibfnamefont {J.}~\bibnamefont {Townsend}}, \bibinfo {author}
  {\bibfnamefont {R.}~\bibnamefont {Christianson}}, \bibinfo {author}
  {\bibfnamefont {Y.~V.}\ \bibnamefont {Tolmachev}}, \bibinfo {author}
  {\bibfnamefont {P.}~\bibnamefont {Keblinski}}, \bibinfo {author}
  {\bibfnamefont {L.-w.}\ \bibnamefont {Hu}}, \bibinfo {author} {\bibfnamefont
  {J.~L.}\ \bibnamefont {Alvarado}},  \emph {et~al.},\ }\bibfield  {title}
  {\enquote {\bibinfo {title} {A benchmark study on the thermal conductivity of
  nanofluids},}\ }\href@noop {} {\bibfield  {journal} {\bibinfo  {journal}
  {Journal of Applied Physics}\ }\textbf {\bibinfo {volume} {106}},\ \bibinfo
  {pages} {094312} (\bibinfo {year} {2009})}\BibitemShut {NoStop}%
\bibitem [{\citenamefont {Lee}\ \emph {et~al.}(2014)\citenamefont {Lee},
  \citenamefont {Alarc\'{o}n-Correa}, \citenamefont {Miksch}, \citenamefont
  {Hahn}, \citenamefont {Gibbs},\ and\ \citenamefont {Fischer}}]{lee2014self}%
  \BibitemOpen
  \bibfield  {author} {\bibinfo {author} {\bibfnamefont {T.-C.}\ \bibnamefont
  {Lee}}, \bibinfo {author} {\bibfnamefont {M.}~\bibnamefont
  {Alarc\'{o}n-Correa}}, \bibinfo {author} {\bibfnamefont {C.}~\bibnamefont
  {Miksch}}, \bibinfo {author} {\bibfnamefont {K.}~\bibnamefont {Hahn}},
  \bibinfo {author} {\bibfnamefont {J.~G.}\ \bibnamefont {Gibbs}}, \ and\
  \bibinfo {author} {\bibfnamefont {P.}~\bibnamefont {Fischer}},\ }\bibfield
  {title} {\enquote {\bibinfo {title} {Self-propelling nanomotors in the
  presence of strong brownian forces},}\ }\href@noop {} {\bibfield  {journal}
  {\bibinfo  {journal} {Nano letters}\ }\textbf {\bibinfo {volume} {14}},\
  \bibinfo {pages} {2407--2412} (\bibinfo {year} {2014})}\BibitemShut {NoStop}%
\bibitem [{\citenamefont {Gao}, \citenamefont {Pei},\ and\ \citenamefont
  {Wang}(2012)}]{gao2012water}%
  \BibitemOpen
  \bibfield  {author} {\bibinfo {author} {\bibfnamefont {W.}~\bibnamefont
  {Gao}}, \bibinfo {author} {\bibfnamefont {A.}~\bibnamefont {Pei}}, \ and\
  \bibinfo {author} {\bibfnamefont {J.}~\bibnamefont {Wang}},\ }\bibfield
  {title} {\enquote {\bibinfo {title} {Water-driven micromotors},}\ }\href@noop
  {} {\bibfield  {journal} {\bibinfo  {journal} {ACS nano}\ }\textbf {\bibinfo
  {volume} {6}},\ \bibinfo {pages} {8432--8438} (\bibinfo {year}
  {2012})}\BibitemShut {NoStop}%
\bibitem [{\citenamefont {Saintillan}(2018)}]{saintillan2018rheology}%
  \BibitemOpen
  \bibfield  {author} {\bibinfo {author} {\bibfnamefont {D.}~\bibnamefont
  {Saintillan}},\ }\bibfield  {title} {\enquote {\bibinfo {title} {Rheology of
  active fluids},}\ }\href@noop {} {\bibfield  {journal} {\bibinfo  {journal}
  {Annual Review of Fluid Mechanics}\ }\textbf {\bibinfo {volume} {50}},\
  \bibinfo {pages} {563--592} (\bibinfo {year} {2018})}\BibitemShut {NoStop}%
\bibitem [{\citenamefont {Guasto}, \citenamefont {Johnson},\ and\ \citenamefont
  {Gollub}(2010)}]{guasto2010oscillatory}%
  \BibitemOpen
  \bibfield  {author} {\bibinfo {author} {\bibfnamefont {J.~S.}\ \bibnamefont
  {Guasto}}, \bibinfo {author} {\bibfnamefont {K.~A.}\ \bibnamefont {Johnson}},
  \ and\ \bibinfo {author} {\bibfnamefont {J.~P.}\ \bibnamefont {Gollub}},\
  }\bibfield  {title} {\enquote {\bibinfo {title} {Oscillatory flows induced by
  microorganisms swimming in two dimensions},}\ }\href@noop {} {\bibfield
  {journal} {\bibinfo  {journal} {Physical review letters}\ }\textbf {\bibinfo
  {volume} {105}},\ \bibinfo {pages} {168102} (\bibinfo {year}
  {2010})}\BibitemShut {NoStop}%
\bibitem [{\citenamefont {Rafa{\"\i}}, \citenamefont {Jibuti},\ and\
  \citenamefont {Peyla}(2010)}]{rafai2010effective}%
  \BibitemOpen
  \bibfield  {author} {\bibinfo {author} {\bibfnamefont {S.}~\bibnamefont
  {Rafa{\"\i}}}, \bibinfo {author} {\bibfnamefont {L.}~\bibnamefont {Jibuti}},
  \ and\ \bibinfo {author} {\bibfnamefont {P.}~\bibnamefont {Peyla}},\
  }\bibfield  {title} {\enquote {\bibinfo {title} {Effective viscosity of
  microswimmer suspensions},}\ }\href@noop {} {\bibfield  {journal} {\bibinfo
  {journal} {Physical Review Letters}\ }\textbf {\bibinfo {volume} {104}},\
  \bibinfo {pages} {098102} (\bibinfo {year} {2010})}\BibitemShut {NoStop}%
\bibitem [{\citenamefont {Wu}\ and\ \citenamefont
  {Libchaber}(2000)}]{wu2000particle}%
  \BibitemOpen
  \bibfield  {author} {\bibinfo {author} {\bibfnamefont {X.-L.}\ \bibnamefont
  {Wu}}\ and\ \bibinfo {author} {\bibfnamefont {A.}~\bibnamefont {Libchaber}},\
  }\bibfield  {title} {\enquote {\bibinfo {title} {Particle diffusion in a
  quasi-two-dimensional bacterial bath},}\ }\href@noop {} {\bibfield  {journal}
  {\bibinfo  {journal} {Physical review letters}\ }\textbf {\bibinfo {volume}
  {84}},\ \bibinfo {pages} {3017} (\bibinfo {year} {2000})}\BibitemShut
  {NoStop}%
\bibitem [{\citenamefont {Kim}\ and\ \citenamefont
  {Breuer}(2004)}]{kim2004enhanced}%
  \BibitemOpen
  \bibfield  {author} {\bibinfo {author} {\bibfnamefont {M.~J.}\ \bibnamefont
  {Kim}}\ and\ \bibinfo {author} {\bibfnamefont {K.~S.}\ \bibnamefont
  {Breuer}},\ }\bibfield  {title} {\enquote {\bibinfo {title} {Enhanced
  diffusion due to motile bacteria},}\ }\href@noop {} {\bibfield  {journal}
  {\bibinfo  {journal} {Physics of fluids}\ }\textbf {\bibinfo {volume} {16}},\
  \bibinfo {pages} {L78--L81} (\bibinfo {year} {2004})}\BibitemShut {NoStop}%
\bibitem [{\citenamefont {Dunkel}\ \emph {et~al.}(2013)\citenamefont {Dunkel},
  \citenamefont {Heidenreich}, \citenamefont {Drescher}, \citenamefont
  {Wensink}, \citenamefont {B{\"a}r},\ and\ \citenamefont
  {Goldstein}}]{dunkel2013fluid}%
  \BibitemOpen
  \bibfield  {author} {\bibinfo {author} {\bibfnamefont {J.}~\bibnamefont
  {Dunkel}}, \bibinfo {author} {\bibfnamefont {S.}~\bibnamefont {Heidenreich}},
  \bibinfo {author} {\bibfnamefont {K.}~\bibnamefont {Drescher}}, \bibinfo
  {author} {\bibfnamefont {H.~H.}\ \bibnamefont {Wensink}}, \bibinfo {author}
  {\bibfnamefont {M.}~\bibnamefont {B{\"a}r}}, \ and\ \bibinfo {author}
  {\bibfnamefont {R.~E.}\ \bibnamefont {Goldstein}},\ }\bibfield  {title}
  {\enquote {\bibinfo {title} {Fluid dynamics of bacterial turbulence},}\
  }\href@noop {} {\bibfield  {journal} {\bibinfo  {journal} {Physical review
  letters}\ }\textbf {\bibinfo {volume} {110}},\ \bibinfo {pages} {228102}
  (\bibinfo {year} {2013})}\BibitemShut {NoStop}%
\bibitem [{\citenamefont {L{\'o}pez}\ \emph {et~al.}(2015)\citenamefont
  {L{\'o}pez}, \citenamefont {Gachelin}, \citenamefont {Douarche},
  \citenamefont {Auradou},\ and\ \citenamefont
  {Cl{\'e}ment}}]{lopez2015turning}%
  \BibitemOpen
  \bibfield  {author} {\bibinfo {author} {\bibfnamefont {H.~M.}\ \bibnamefont
  {L{\'o}pez}}, \bibinfo {author} {\bibfnamefont {J.}~\bibnamefont {Gachelin}},
  \bibinfo {author} {\bibfnamefont {C.}~\bibnamefont {Douarche}}, \bibinfo
  {author} {\bibfnamefont {H.}~\bibnamefont {Auradou}}, \ and\ \bibinfo
  {author} {\bibfnamefont {E.}~\bibnamefont {Cl{\'e}ment}},\ }\bibfield
  {title} {\enquote {\bibinfo {title} {Turning bacteria suspensions into
  superfluids},}\ }\href@noop {} {\bibfield  {journal} {\bibinfo  {journal}
  {Physical review letters}\ }\textbf {\bibinfo {volume} {115}},\ \bibinfo
  {pages} {028301} (\bibinfo {year} {2015})}\BibitemShut {NoStop}%
\bibitem [{\citenamefont {Saintillan}\ and\ \citenamefont
  {Shelley}(2012)}]{saintillan2012emergence}%
  \BibitemOpen
  \bibfield  {author} {\bibinfo {author} {\bibfnamefont {D.}~\bibnamefont
  {Saintillan}}\ and\ \bibinfo {author} {\bibfnamefont {M.~J.}\ \bibnamefont
  {Shelley}},\ }\bibfield  {title} {\enquote {\bibinfo {title} {Emergence of
  coherent structures and large-scale flows in motile suspensions},}\
  }\href@noop {} {\bibfield  {journal} {\bibinfo  {journal} {Journal of the
  Royal Society Interface}\ }\textbf {\bibinfo {volume} {9}},\ \bibinfo {pages}
  {571--585} (\bibinfo {year} {2012})}\BibitemShut {NoStop}%
\bibitem [{\citenamefont {Nishiguchi}\ and\ \citenamefont
  {Sano}(2015)}]{nishiguchi2015mesoscopic}%
  \BibitemOpen
  \bibfield  {author} {\bibinfo {author} {\bibfnamefont {D.}~\bibnamefont
  {Nishiguchi}}\ and\ \bibinfo {author} {\bibfnamefont {M.}~\bibnamefont
  {Sano}},\ }\bibfield  {title} {\enquote {\bibinfo {title} {Mesoscopic
  turbulence and local order in janus particles self-propelling under an ac
  electric field},}\ }\href@noop {} {\bibfield  {journal} {\bibinfo  {journal}
  {Physical Review E}\ }\textbf {\bibinfo {volume} {92}},\ \bibinfo {pages}
  {052309} (\bibinfo {year} {2015})}\BibitemShut {NoStop}%
\bibitem [{\citenamefont {Gregory}\ and\ \citenamefont
  {Ebbens}(2018)}]{gregory2018symmetrical}%
  \BibitemOpen
  \bibfield  {author} {\bibinfo {author} {\bibfnamefont {D.~A.}\ \bibnamefont
  {Gregory}}\ and\ \bibinfo {author} {\bibfnamefont {S.~J.}\ \bibnamefont
  {Ebbens}},\ }\bibfield  {title} {\enquote {\bibinfo {title} {Symmetrical
  catalytically active colloids collectively induce convective flow},}\
  }\href@noop {} {\bibfield  {journal} {\bibinfo  {journal} {Langmuir}\
  }\textbf {\bibinfo {volume} {34}},\ \bibinfo {pages} {4307--4313} (\bibinfo
  {year} {2018})}\BibitemShut {NoStop}%
\bibitem [{\citenamefont {El~Hasadi}\ and\ \citenamefont
  {Crapper}(2017)}]{el2017self}%
  \BibitemOpen
  \bibfield  {author} {\bibinfo {author} {\bibfnamefont {Y.~M.}\ \bibnamefont
  {El~Hasadi}}\ and\ \bibinfo {author} {\bibfnamefont {M.}~\bibnamefont
  {Crapper}},\ }\bibfield  {title} {\enquote {\bibinfo {title} {Self-propelled
  nanofluids a path to a highly effective coolant},}\ }\href@noop {} {\bibfield
   {journal} {\bibinfo  {journal} {Applied Thermal Engineering}\ }\textbf
  {\bibinfo {volume} {127}},\ \bibinfo {pages} {857--869} (\bibinfo {year}
  {2017})}\BibitemShut {NoStop}%
\bibitem [{\citenamefont {El~Hasadi}\ and\ \citenamefont
  {Crapper}(2020)}]{el2020self}%
  \BibitemOpen
  \bibfield  {author} {\bibinfo {author} {\bibfnamefont {Y.~M.}\ \bibnamefont
  {El~Hasadi}}\ and\ \bibinfo {author} {\bibfnamefont {M.}~\bibnamefont
  {Crapper}},\ }\bibfield  {title} {\enquote {\bibinfo {title} {Self-propelled
  nanofluids a coolant inspired from nature with enhanced thermal transport
  properties},}\ }\href@noop {} {\bibfield  {journal} {\bibinfo  {journal}
  {Journal of Molecular Liquids}\ ,\ \bibinfo {pages} {113548}} (\bibinfo
  {year} {2020})}\BibitemShut {NoStop}%
\bibitem [{\citenamefont {Gachelin}\ \emph {et~al.}(2013)\citenamefont
  {Gachelin}, \citenamefont {Mino}, \citenamefont {Berthet}, \citenamefont
  {Lindner}, \citenamefont {Rousselet},\ and\ \citenamefont
  {Cl{\'e}ment}}]{gachelin2013non}%
  \BibitemOpen
  \bibfield  {author} {\bibinfo {author} {\bibfnamefont {J.}~\bibnamefont
  {Gachelin}}, \bibinfo {author} {\bibfnamefont {G.}~\bibnamefont {Mino}},
  \bibinfo {author} {\bibfnamefont {H.}~\bibnamefont {Berthet}}, \bibinfo
  {author} {\bibfnamefont {A.}~\bibnamefont {Lindner}}, \bibinfo {author}
  {\bibfnamefont {A.}~\bibnamefont {Rousselet}}, \ and\ \bibinfo {author}
  {\bibfnamefont {{\'E}.}~\bibnamefont {Cl{\'e}ment}},\ }\bibfield  {title}
  {\enquote {\bibinfo {title} {Non-newtonian viscosity of escherichia coli
  suspensions},}\ }\href@noop {} {\bibfield  {journal} {\bibinfo  {journal}
  {Physical review letters}\ }\textbf {\bibinfo {volume} {110}},\ \bibinfo
  {pages} {268103} (\bibinfo {year} {2013})}\BibitemShut {NoStop}%
\bibitem [{\citenamefont {Ghosh}\ and\ \citenamefont
  {Fischer}(2009)}]{ghosh2009controlled}%
  \BibitemOpen
  \bibfield  {author} {\bibinfo {author} {\bibfnamefont {A.}~\bibnamefont
  {Ghosh}}\ and\ \bibinfo {author} {\bibfnamefont {P.}~\bibnamefont
  {Fischer}},\ }\bibfield  {title} {\enquote {\bibinfo {title} {Controlled
  propulsion of artificial magnetic nanostructured propellers},}\ }\href@noop
  {} {\bibfield  {journal} {\bibinfo  {journal} {Nano letters}\ }\textbf
  {\bibinfo {volume} {9}},\ \bibinfo {pages} {2243--2245} (\bibinfo {year}
  {2009})}\BibitemShut {NoStop}%
\bibitem [{\citenamefont {Schamel}\ \emph {et~al.}(2014)\citenamefont
  {Schamel}, \citenamefont {Mark}, \citenamefont {Gibbs}, \citenamefont
  {Miksch}, \citenamefont {Morozov}, \citenamefont {Leshansky},\ and\
  \citenamefont {Fischer}}]{schamel2014nanopropellers}%
  \BibitemOpen
  \bibfield  {author} {\bibinfo {author} {\bibfnamefont {D.}~\bibnamefont
  {Schamel}}, \bibinfo {author} {\bibfnamefont {A.~G.}\ \bibnamefont {Mark}},
  \bibinfo {author} {\bibfnamefont {J.~G.}\ \bibnamefont {Gibbs}}, \bibinfo
  {author} {\bibfnamefont {C.}~\bibnamefont {Miksch}}, \bibinfo {author}
  {\bibfnamefont {K.~I.}\ \bibnamefont {Morozov}}, \bibinfo {author}
  {\bibfnamefont {A.~M.}\ \bibnamefont {Leshansky}}, \ and\ \bibinfo {author}
  {\bibfnamefont {P.}~\bibnamefont {Fischer}},\ }\bibfield  {title} {\enquote
  {\bibinfo {title} {Nanopropellers and their actuation in complex viscoelastic
  media},}\ }\href@noop {} {\bibfield  {journal} {\bibinfo  {journal} {ACS
  nano}\ }\textbf {\bibinfo {volume} {8}},\ \bibinfo {pages} {8794--8801}
  (\bibinfo {year} {2014})}\BibitemShut {NoStop}%
\bibitem [{\citenamefont {Kremer}\ and\ \citenamefont
  {Grest}(1990)}]{kremer1990dynamics}%
  \BibitemOpen
  \bibfield  {author} {\bibinfo {author} {\bibfnamefont {K.}~\bibnamefont
  {Kremer}}\ and\ \bibinfo {author} {\bibfnamefont {G.~S.}\ \bibnamefont
  {Grest}},\ }\bibfield  {title} {\enquote {\bibinfo {title} {Dynamics of
  entangled linear polymer melts: A molecular-dynamics simulation},}\
  }\href@noop {} {\bibfield  {journal} {\bibinfo  {journal} {The Journal of
  Chemical Physics}\ }\textbf {\bibinfo {volume} {92}},\ \bibinfo {pages}
  {5057--5086} (\bibinfo {year} {1990})}\BibitemShut {NoStop}%
\bibitem [{\citenamefont {Andersen}, \citenamefont {Weeks},\ and\ \citenamefont
  {Chandler}(1971)}]{andersen1971relationship}%
  \BibitemOpen
  \bibfield  {author} {\bibinfo {author} {\bibfnamefont {H.~C.}\ \bibnamefont
  {Andersen}}, \bibinfo {author} {\bibfnamefont {J.~D.}\ \bibnamefont {Weeks}},
  \ and\ \bibinfo {author} {\bibfnamefont {D.}~\bibnamefont {Chandler}},\
  }\bibfield  {title} {\enquote {\bibinfo {title} {Relationship between the
  hard-sphere fluid and fluids with realistic repulsive forces},}\ }\href@noop
  {} {\bibfield  {journal} {\bibinfo  {journal} {Physical Review A}\ }\textbf
  {\bibinfo {volume} {4}},\ \bibinfo {pages} {1597} (\bibinfo {year}
  {1971})}\BibitemShut {NoStop}%
\bibitem [{\citenamefont {Plimpton}(1995)}]{plimpton1995fast}%
  \BibitemOpen
  \bibfield  {author} {\bibinfo {author} {\bibfnamefont {S.}~\bibnamefont
  {Plimpton}},\ }\bibfield  {title} {\enquote {\bibinfo {title} {Fast parallel
  algorithms for short-range molecular dynamics},}\ }\href@noop {} {\bibfield
  {journal} {\bibinfo  {journal} {Journal of computational physics}\ }\textbf
  {\bibinfo {volume} {117}},\ \bibinfo {pages} {1--19} (\bibinfo {year}
  {1995})}\BibitemShut {NoStop}%
\bibitem [{\citenamefont {Martyna}, \citenamefont {Tobias},\ and\ \citenamefont
  {Klein}(1994)}]{martyna1994constant}%
  \BibitemOpen
  \bibfield  {author} {\bibinfo {author} {\bibfnamefont {G.~J.}\ \bibnamefont
  {Martyna}}, \bibinfo {author} {\bibfnamefont {D.~J.}\ \bibnamefont {Tobias}},
  \ and\ \bibinfo {author} {\bibfnamefont {M.~L.}\ \bibnamefont {Klein}},\
  }\bibfield  {title} {\enquote {\bibinfo {title} {Constant pressure molecular
  dynamics algorithms},}\ }\href@noop {} {\bibfield  {journal} {\bibinfo
  {journal} {The Journal of chemical physics}\ }\textbf {\bibinfo {volume}
  {101}},\ \bibinfo {pages} {4177--4189} (\bibinfo {year} {1994})}\BibitemShut
  {NoStop}%
\bibitem [{\citenamefont {Nos{\'e}}(1984)}]{nose1984unified}%
  \BibitemOpen
  \bibfield  {author} {\bibinfo {author} {\bibfnamefont {S.}~\bibnamefont
  {Nos{\'e}}},\ }\bibfield  {title} {\enquote {\bibinfo {title} {A unified
  formulation of the constant temperature molecular dynamics methods},}\
  }\href@noop {} {\bibfield  {journal} {\bibinfo  {journal} {The Journal of
  chemical physics}\ }\textbf {\bibinfo {volume} {81}},\ \bibinfo {pages}
  {511--519} (\bibinfo {year} {1984})}\BibitemShut {NoStop}%
\bibitem [{\citenamefont {Hoover}(1985)}]{hoover1985canonical}%
  \BibitemOpen
  \bibfield  {author} {\bibinfo {author} {\bibfnamefont {W.~G.}\ \bibnamefont
  {Hoover}},\ }\bibfield  {title} {\enquote {\bibinfo {title} {Canonical
  dynamics: Equilibrium phase-space distributions},}\ }\href@noop {} {\bibfield
   {journal} {\bibinfo  {journal} {Physical review A}\ }\textbf {\bibinfo
  {volume} {31}},\ \bibinfo {pages} {1695} (\bibinfo {year}
  {1985})}\BibitemShut {NoStop}%
\bibitem [{\citenamefont {Purcell}(1977)}]{purcell1977life}%
  \BibitemOpen
  \bibfield  {author} {\bibinfo {author} {\bibfnamefont {E.~M.}\ \bibnamefont
  {Purcell}},\ }\bibfield  {title} {\enquote {\bibinfo {title} {Life at low
  reynolds number},}\ }\href@noop {} {\bibfield  {journal} {\bibinfo  {journal}
  {American journal of physics}\ }\textbf {\bibinfo {volume} {45}},\ \bibinfo
  {pages} {3--11} (\bibinfo {year} {1977})}\BibitemShut {NoStop}%
\bibitem [{\citenamefont {Ten~Hagen}\ \emph {et~al.}(2015)\citenamefont
  {Ten~Hagen}, \citenamefont {Wittkowski}, \citenamefont {Takagi},
  \citenamefont {K{\"u}mmel}, \citenamefont {Bechinger},\ and\ \citenamefont
  {L{\"o}wen}}]{ten2015can}%
  \BibitemOpen
  \bibfield  {author} {\bibinfo {author} {\bibfnamefont {B.}~\bibnamefont
  {Ten~Hagen}}, \bibinfo {author} {\bibfnamefont {R.}~\bibnamefont
  {Wittkowski}}, \bibinfo {author} {\bibfnamefont {D.}~\bibnamefont {Takagi}},
  \bibinfo {author} {\bibfnamefont {F.}~\bibnamefont {K{\"u}mmel}}, \bibinfo
  {author} {\bibfnamefont {C.}~\bibnamefont {Bechinger}}, \ and\ \bibinfo
  {author} {\bibfnamefont {H.}~\bibnamefont {L{\"o}wen}},\ }\bibfield  {title}
  {\enquote {\bibinfo {title} {Can the self-propulsion of anisotropic
  microswimmers be described by using forces and torques?}}\ }\href@noop {}
  {\bibfield  {journal} {\bibinfo  {journal} {Journal of Physics: Condensed
  Matter}\ }\textbf {\bibinfo {volume} {27}},\ \bibinfo {pages} {194110}
  (\bibinfo {year} {2015})}\BibitemShut {NoStop}%
\bibitem [{\citenamefont {Moran}\ and\ \citenamefont
  {Posner}(2017)}]{moran2017phoretic}%
  \BibitemOpen
  \bibfield  {author} {\bibinfo {author} {\bibfnamefont {J.~L.}\ \bibnamefont
  {Moran}}\ and\ \bibinfo {author} {\bibfnamefont {J.~D.}\ \bibnamefont
  {Posner}},\ }\bibfield  {title} {\enquote {\bibinfo {title} {Phoretic
  self-propulsion},}\ }\href@noop {} {\bibfield  {journal} {\bibinfo  {journal}
  {Annual Review of Fluid Mechanics}\ }\textbf {\bibinfo {volume} {49}},\
  \bibinfo {pages} {511--540} (\bibinfo {year} {2017})}\BibitemShut {NoStop}%
\bibitem [{\citenamefont {Golestanian}, \citenamefont {Liverpool},\ and\
  \citenamefont {Ajdari}(2007)}]{golestanian2007designing}%
  \BibitemOpen
  \bibfield  {author} {\bibinfo {author} {\bibfnamefont {R.}~\bibnamefont
  {Golestanian}}, \bibinfo {author} {\bibfnamefont {T.}~\bibnamefont
  {Liverpool}}, \ and\ \bibinfo {author} {\bibfnamefont {A.}~\bibnamefont
  {Ajdari}},\ }\bibfield  {title} {\enquote {\bibinfo {title} {Designing
  phoretic micro-and nano-swimmers},}\ }\href@noop {} {\bibfield  {journal}
  {\bibinfo  {journal} {New Journal of Physics}\ }\textbf {\bibinfo {volume}
  {9}},\ \bibinfo {pages} {126} (\bibinfo {year} {2007})}\BibitemShut {NoStop}%
\bibitem [{\citenamefont {Happel}\ and\ \citenamefont
  {Brenner}(2012)}]{happel2012low}%
  \BibitemOpen
  \bibfield  {author} {\bibinfo {author} {\bibfnamefont {J.}~\bibnamefont
  {Happel}}\ and\ \bibinfo {author} {\bibfnamefont {H.}~\bibnamefont
  {Brenner}},\ }\href@noop {} {\emph {\bibinfo {title} {Low Reynolds number
  hydrodynamics: with special applications to particulate media}}},\
  Vol.~\bibinfo {volume} {1}\ (\bibinfo  {publisher} {Springer Science \&
  Business Media},\ \bibinfo {year} {2012})\BibitemShut {NoStop}%
\bibitem [{\citenamefont {Stone}\ and\ \citenamefont
  {Samuel}(1996)}]{stone1996propulsion}%
  \BibitemOpen
  \bibfield  {author} {\bibinfo {author} {\bibfnamefont {H.~A.}\ \bibnamefont
  {Stone}}\ and\ \bibinfo {author} {\bibfnamefont {A.~D.}\ \bibnamefont
  {Samuel}},\ }\bibfield  {title} {\enquote {\bibinfo {title} {Propulsion of
  microorganisms by surface distortions},}\ }\href@noop {} {\bibfield
  {journal} {\bibinfo  {journal} {Physical review letters}\ }\textbf {\bibinfo
  {volume} {77}},\ \bibinfo {pages} {4102} (\bibinfo {year}
  {1996})}\BibitemShut {NoStop}%
\bibitem [{\citenamefont {Lauga}(2020)}]{lauga2020fluid}%
  \BibitemOpen
  \bibfield  {author} {\bibinfo {author} {\bibfnamefont {E.}~\bibnamefont
  {Lauga}},\ }\href@noop {} {\emph {\bibinfo {title} {The Fluid Dynamics of
  Cell Motility}}},\ Vol.~\bibinfo {volume} {62}\ (\bibinfo  {publisher}
  {Cambridge University Press},\ \bibinfo {year} {2020})\BibitemShut {NoStop}%
\bibitem [{\citenamefont {Paxton}\ \emph {et~al.}(2006)\citenamefont {Paxton},
  \citenamefont {Baker}, \citenamefont {Kline}, \citenamefont {Wang},
  \citenamefont {Mallouk},\ and\ \citenamefont
  {Sen}}]{paxton2006catalytically}%
  \BibitemOpen
  \bibfield  {author} {\bibinfo {author} {\bibfnamefont {W.~F.}\ \bibnamefont
  {Paxton}}, \bibinfo {author} {\bibfnamefont {P.~T.}\ \bibnamefont {Baker}},
  \bibinfo {author} {\bibfnamefont {T.~R.}\ \bibnamefont {Kline}}, \bibinfo
  {author} {\bibfnamefont {Y.}~\bibnamefont {Wang}}, \bibinfo {author}
  {\bibfnamefont {T.~E.}\ \bibnamefont {Mallouk}}, \ and\ \bibinfo {author}
  {\bibfnamefont {A.}~\bibnamefont {Sen}},\ }\bibfield  {title} {\enquote
  {\bibinfo {title} {Catalytically induced electrokinetics for motors and
  micropumps},}\ }\href@noop {} {\bibfield  {journal} {\bibinfo  {journal}
  {Journal of the American Chemical Society}\ }\textbf {\bibinfo {volume}
  {128}},\ \bibinfo {pages} {14881--14888} (\bibinfo {year}
  {2006})}\BibitemShut {NoStop}%
\bibitem [{\citenamefont {Moran}\ and\ \citenamefont
  {Posner}(2014)}]{moran2014role}%
  \BibitemOpen
  \bibfield  {author} {\bibinfo {author} {\bibfnamefont {J.~L.}\ \bibnamefont
  {Moran}}\ and\ \bibinfo {author} {\bibfnamefont {J.~D.}\ \bibnamefont
  {Posner}},\ }\bibfield  {title} {\enquote {\bibinfo {title} {Role of solution
  conductivity in reaction induced charge auto-electrophoresis},}\ }\href@noop
  {} {\bibfield  {journal} {\bibinfo  {journal} {Physics of Fluids}\ }\textbf
  {\bibinfo {volume} {26}},\ \bibinfo {pages} {042001} (\bibinfo {year}
  {2014})}\BibitemShut {NoStop}%
\bibitem [{\citenamefont {Guzm{\'a}n-Lastra}, \citenamefont {Kaiser},\ and\
  \citenamefont {L{\"o}wen}(2016)}]{guzman2016fission}%
  \BibitemOpen
  \bibfield  {author} {\bibinfo {author} {\bibfnamefont {F.}~\bibnamefont
  {Guzm{\'a}n-Lastra}}, \bibinfo {author} {\bibfnamefont {A.}~\bibnamefont
  {Kaiser}}, \ and\ \bibinfo {author} {\bibfnamefont {H.}~\bibnamefont
  {L{\"o}wen}},\ }\bibfield  {title} {\enquote {\bibinfo {title} {Fission and
  fusion scenarios for magnetic microswimmer clusters},}\ }\href@noop {}
  {\bibfield  {journal} {\bibinfo  {journal} {Nature communications}\ }\textbf
  {\bibinfo {volume} {7}},\ \bibinfo {pages} {1--11} (\bibinfo {year}
  {2016})}\BibitemShut {NoStop}%
\bibitem [{\citenamefont {Marchetti}\ \emph {et~al.}(2013)\citenamefont
  {Marchetti}, \citenamefont {Joanny}, \citenamefont {Ramaswamy}, \citenamefont
  {Liverpool}, \citenamefont {Prost}, \citenamefont {Rao},\ and\ \citenamefont
  {Simha}}]{marchetti2013hydrodynamics}%
  \BibitemOpen
  \bibfield  {author} {\bibinfo {author} {\bibfnamefont {M.~C.}\ \bibnamefont
  {Marchetti}}, \bibinfo {author} {\bibfnamefont {J.-F.}\ \bibnamefont
  {Joanny}}, \bibinfo {author} {\bibfnamefont {S.}~\bibnamefont {Ramaswamy}},
  \bibinfo {author} {\bibfnamefont {T.~B.}\ \bibnamefont {Liverpool}}, \bibinfo
  {author} {\bibfnamefont {J.}~\bibnamefont {Prost}}, \bibinfo {author}
  {\bibfnamefont {M.}~\bibnamefont {Rao}}, \ and\ \bibinfo {author}
  {\bibfnamefont {R.~A.}\ \bibnamefont {Simha}},\ }\bibfield  {title} {\enquote
  {\bibinfo {title} {Hydrodynamics of soft active matter},}\ }\href@noop {}
  {\bibfield  {journal} {\bibinfo  {journal} {Reviews of Modern Physics}\
  }\textbf {\bibinfo {volume} {85}},\ \bibinfo {pages} {1143} (\bibinfo {year}
  {2013})}\BibitemShut {NoStop}%
\bibitem [{\citenamefont {Elgeti}, \citenamefont {Winkler},\ and\ \citenamefont
  {Gompper}(2015)}]{elgeti2015physics}%
  \BibitemOpen
  \bibfield  {author} {\bibinfo {author} {\bibfnamefont {J.}~\bibnamefont
  {Elgeti}}, \bibinfo {author} {\bibfnamefont {R.~G.}\ \bibnamefont {Winkler}},
  \ and\ \bibinfo {author} {\bibfnamefont {G.}~\bibnamefont {Gompper}},\
  }\bibfield  {title} {\enquote {\bibinfo {title} {Physics of
  microswimmers—single particle motion and collective behavior: a review},}\
  }\href@noop {} {\bibfield  {journal} {\bibinfo  {journal} {Reports on
  progress in physics}\ }\textbf {\bibinfo {volume} {78}},\ \bibinfo {pages}
  {056601} (\bibinfo {year} {2015})}\BibitemShut {NoStop}%
\bibitem [{\citenamefont {Green}(1954)}]{green1954markoff}%
  \BibitemOpen
  \bibfield  {author} {\bibinfo {author} {\bibfnamefont {M.~S.}\ \bibnamefont
  {Green}},\ }\bibfield  {title} {\enquote {\bibinfo {title} {Markoff random
  processes and the statistical mechanics of time-dependent phenomena. ii.
  irreversible processes in fluids},}\ }\href@noop {} {\bibfield  {journal}
  {\bibinfo  {journal} {The Journal of Chemical Physics}\ }\textbf {\bibinfo
  {volume} {22}},\ \bibinfo {pages} {398--413} (\bibinfo {year}
  {1954})}\BibitemShut {NoStop}%
\bibitem [{\citenamefont {Kubo}(1957)}]{kubo1957statistical}%
  \BibitemOpen
  \bibfield  {author} {\bibinfo {author} {\bibfnamefont {R.}~\bibnamefont
  {Kubo}},\ }\bibfield  {title} {\enquote {\bibinfo {title}
  {Statistical-mechanical theory of irreversible processes. i. general theory
  and simple applications to magnetic and conduction problems},}\ }\href@noop
  {} {\bibfield  {journal} {\bibinfo  {journal} {Journal of the Physical
  Society of Japan}\ }\textbf {\bibinfo {volume} {12}},\ \bibinfo {pages}
  {570--586} (\bibinfo {year} {1957})}\BibitemShut {NoStop}%
\bibitem [{\citenamefont {Asta}\ \emph {et~al.}(2017)\citenamefont {Asta},
  \citenamefont {Levesque}, \citenamefont {Vuilleumier},\ and\ \citenamefont
  {Rotenberg}}]{asta2017transient}%
  \BibitemOpen
  \bibfield  {author} {\bibinfo {author} {\bibfnamefont {A.~J.}\ \bibnamefont
  {Asta}}, \bibinfo {author} {\bibfnamefont {M.}~\bibnamefont {Levesque}},
  \bibinfo {author} {\bibfnamefont {R.}~\bibnamefont {Vuilleumier}}, \ and\
  \bibinfo {author} {\bibfnamefont {B.}~\bibnamefont {Rotenberg}},\ }\bibfield
  {title} {\enquote {\bibinfo {title} {Transient hydrodynamic finite-size
  effects in simulations under periodic boundary conditions},}\ }\href@noop {}
  {\bibfield  {journal} {\bibinfo  {journal} {Physical Review E}\ }\textbf
  {\bibinfo {volume} {95}},\ \bibinfo {pages} {061301} (\bibinfo {year}
  {2017})}\BibitemShut {NoStop}%
\bibitem [{\citenamefont {Romanczuk}\ \emph {et~al.}(2012)\citenamefont
  {Romanczuk}, \citenamefont {B{\"a}r}, \citenamefont {Ebeling}, \citenamefont
  {Lindner},\ and\ \citenamefont {Schimansky-Geier}}]{romanczuk2012active}%
  \BibitemOpen
  \bibfield  {author} {\bibinfo {author} {\bibfnamefont {P.}~\bibnamefont
  {Romanczuk}}, \bibinfo {author} {\bibfnamefont {M.}~\bibnamefont {B{\"a}r}},
  \bibinfo {author} {\bibfnamefont {W.}~\bibnamefont {Ebeling}}, \bibinfo
  {author} {\bibfnamefont {B.}~\bibnamefont {Lindner}}, \ and\ \bibinfo
  {author} {\bibfnamefont {L.}~\bibnamefont {Schimansky-Geier}},\ }\bibfield
  {title} {\enquote {\bibinfo {title} {Active brownian particles},}\
  }\href@noop {} {\bibfield  {journal} {\bibinfo  {journal} {The European
  Physical Journal Special Topics}\ }\textbf {\bibinfo {volume} {202}},\
  \bibinfo {pages} {1--162} (\bibinfo {year} {2012})}\BibitemShut {NoStop}%
\bibitem [{\citenamefont {Takatori}\ and\ \citenamefont
  {Brady}(2016)}]{takatori2016forces}%
  \BibitemOpen
  \bibfield  {author} {\bibinfo {author} {\bibfnamefont {S.~C.}\ \bibnamefont
  {Takatori}}\ and\ \bibinfo {author} {\bibfnamefont {J.~F.}\ \bibnamefont
  {Brady}},\ }\bibfield  {title} {\enquote {\bibinfo {title} {Forces, stresses
  and the (thermo?) dynamics of active matter},}\ }\href@noop {} {\bibfield
  {journal} {\bibinfo  {journal} {Current Opinion in Colloid \& Interface
  Science}\ }\textbf {\bibinfo {volume} {21}},\ \bibinfo {pages} {24--33}
  (\bibinfo {year} {2016})}\BibitemShut {NoStop}%
\bibitem [{\citenamefont {Ohtori}, \citenamefont {Uchiyama},\ and\
  \citenamefont {Ishii}(2018)}]{ohtori2018stokes}%
  \BibitemOpen
  \bibfield  {author} {\bibinfo {author} {\bibfnamefont {N.}~\bibnamefont
  {Ohtori}}, \bibinfo {author} {\bibfnamefont {H.}~\bibnamefont {Uchiyama}}, \
  and\ \bibinfo {author} {\bibfnamefont {Y.}~\bibnamefont {Ishii}},\ }\bibfield
   {title} {\enquote {\bibinfo {title} {The stokes-einstein relation for simple
  fluids: From hard-sphere to lennard-jones via wca potentials},}\ }\href@noop
  {} {\bibfield  {journal} {\bibinfo  {journal} {The Journal of chemical
  physics}\ }\textbf {\bibinfo {volume} {149}},\ \bibinfo {pages} {214501}
  (\bibinfo {year} {2018})}\BibitemShut {NoStop}%
\bibitem [{\citenamefont {Howse}\ \emph {et~al.}(2007)\citenamefont {Howse},
  \citenamefont {Jones}, \citenamefont {Ryan}, \citenamefont {Gough},
  \citenamefont {Vafabakhsh},\ and\ \citenamefont
  {Golestanian}}]{howse2007self}%
  \BibitemOpen
  \bibfield  {author} {\bibinfo {author} {\bibfnamefont {J.~R.}\ \bibnamefont
  {Howse}}, \bibinfo {author} {\bibfnamefont {R.~A.}\ \bibnamefont {Jones}},
  \bibinfo {author} {\bibfnamefont {A.~J.}\ \bibnamefont {Ryan}}, \bibinfo
  {author} {\bibfnamefont {T.}~\bibnamefont {Gough}}, \bibinfo {author}
  {\bibfnamefont {R.}~\bibnamefont {Vafabakhsh}}, \ and\ \bibinfo {author}
  {\bibfnamefont {R.}~\bibnamefont {Golestanian}},\ }\bibfield  {title}
  {\enquote {\bibinfo {title} {Self-motile colloidal particles: from directed
  propulsion to random walk},}\ }\href@noop {} {\bibfield  {journal} {\bibinfo
  {journal} {Physical review letters}\ }\textbf {\bibinfo {volume} {99}},\
  \bibinfo {pages} {048102} (\bibinfo {year} {2007})}\BibitemShut {NoStop}%
\bibitem [{\citenamefont {Bechinger}\ \emph {et~al.}(2016)\citenamefont
  {Bechinger}, \citenamefont {Di~Leonardo}, \citenamefont {L{\"o}wen},
  \citenamefont {Reichhardt}, \citenamefont {Volpe},\ and\ \citenamefont
  {Volpe}}]{bechinger2016active}%
  \BibitemOpen
  \bibfield  {author} {\bibinfo {author} {\bibfnamefont {C.}~\bibnamefont
  {Bechinger}}, \bibinfo {author} {\bibfnamefont {R.}~\bibnamefont
  {Di~Leonardo}}, \bibinfo {author} {\bibfnamefont {H.}~\bibnamefont
  {L{\"o}wen}}, \bibinfo {author} {\bibfnamefont {C.}~\bibnamefont
  {Reichhardt}}, \bibinfo {author} {\bibfnamefont {G.}~\bibnamefont {Volpe}}, \
  and\ \bibinfo {author} {\bibfnamefont {G.}~\bibnamefont {Volpe}},\ }\bibfield
   {title} {\enquote {\bibinfo {title} {Active particles in complex and crowded
  environments},}\ }\href@noop {} {\bibfield  {journal} {\bibinfo  {journal}
  {Reviews of Modern Physics}\ }\textbf {\bibinfo {volume} {88}},\ \bibinfo
  {pages} {045006} (\bibinfo {year} {2016})}\BibitemShut {NoStop}%
\bibitem [{\citenamefont {Thampi}\ and\ \citenamefont
  {Yeomans}(2016)}]{thampi2016active}%
  \BibitemOpen
  \bibfield  {author} {\bibinfo {author} {\bibfnamefont {S.}~\bibnamefont
  {Thampi}}\ and\ \bibinfo {author} {\bibfnamefont {J.}~\bibnamefont
  {Yeomans}},\ }\bibfield  {title} {\enquote {\bibinfo {title} {Active
  turbulence in active nematics},}\ }\href@noop {} {\bibfield  {journal}
  {\bibinfo  {journal} {The European Physical Journal Special Topics}\ }\textbf
  {\bibinfo {volume} {225}},\ \bibinfo {pages} {651--662} (\bibinfo {year}
  {2016})}\BibitemShut {NoStop}%
\bibitem [{\citenamefont {Chiang}\ and\ \citenamefont
  {Velegol}(2014)}]{chiang2014localized}%
  \BibitemOpen
  \bibfield  {author} {\bibinfo {author} {\bibfnamefont {T.-Y.}\ \bibnamefont
  {Chiang}}\ and\ \bibinfo {author} {\bibfnamefont {D.}~\bibnamefont
  {Velegol}},\ }\bibfield  {title} {\enquote {\bibinfo {title} {Localized
  electroosmosis (leo) induced by spherical colloidal motors},}\ }\href@noop {}
  {\bibfield  {journal} {\bibinfo  {journal} {Langmuir}\ }\textbf {\bibinfo
  {volume} {30}},\ \bibinfo {pages} {2600--2607} (\bibinfo {year}
  {2014})}\BibitemShut {NoStop}%
\bibitem [{\citenamefont {Campbell}\ \emph {et~al.}(2019)\citenamefont
  {Campbell}, \citenamefont {Ebbens}, \citenamefont {Illien},\ and\
  \citenamefont {Golestanian}}]{campbell2019experimental}%
  \BibitemOpen
  \bibfield  {author} {\bibinfo {author} {\bibfnamefont {A.~I.}\ \bibnamefont
  {Campbell}}, \bibinfo {author} {\bibfnamefont {S.~J.}\ \bibnamefont
  {Ebbens}}, \bibinfo {author} {\bibfnamefont {P.}~\bibnamefont {Illien}}, \
  and\ \bibinfo {author} {\bibfnamefont {R.}~\bibnamefont {Golestanian}},\
  }\bibfield  {title} {\enquote {\bibinfo {title} {Experimental observation of
  flow fields around active janus spheres},}\ }\href@noop {} {\bibfield
  {journal} {\bibinfo  {journal} {Nature communications}\ }\textbf {\bibinfo
  {volume} {10}},\ \bibinfo {pages} {1--8} (\bibinfo {year}
  {2019})}\BibitemShut {NoStop}%
\end{thebibliography}%

\end{document}